\documentclass[11pt]{amsart}
\usepackage{stmaryrd}
\usepackage{color}
\usepackage{amsmath}
\usepackage{amssymb}
\usepackage{float}
\usepackage{graphicx}

\newtheorem{example}{Example}[section]

\newtheorem{note}{Remark}[section]

\newsavebox{\savepar}

\def\nn{\mathbb{N}}

\def\zz{\mathbb{Z}}

\def\F{\mathcal{F}}

\def\epsilon{\varepsilon}
\def\leq{\leqslant}
\def\geq{\geqslant}

\def\x{{\bf x}}

\def\ms{{\medskip\noindent}}
\def\bs{{\bigskip\noindent}}

\def\int{{\rm int}}

\def\Vext{{V_{\rm ext}}}

\begin{document}
\title{Open Regulatory Networks and Modularity}

\author{R. Lima, A. Meyroneinc \& E. Ugalde}

\maketitle

\vskip 1truecm

\begin{center}
\begin{minipage}{13truecm}
\begin{small}
Centre de Physique Th\'eorique, CNRS--Luminy Case 907, 
13288 Marseille Cedex 09, France, lima@cpt.univ-mrs.fr\\
Departamento de Matem\'aticas, Instituto Venezolano de 
Investigaciones Cient\'\i ficas, Apartado 21827, Caracas 1020A, 
Venezuela, ameyrone@ivic.ve\\
Instituto de F\'\i sica, Universidad Aut\'onoma de San Luis Potos\'\i , 
Av. Manuel Nava 6, San Luis Potos\'\i , 78290 M\'exico, ugalde@ifisica.uaslp.mx
\end{small}
\end{minipage}
\end{center}

\vskip 3truecm

\begin{abstract}
We study the dynamical properties of small regulatory networks treated as non 
autonomous dynamical systems called \emph{modules} when working inside larger 
networks or, equivalently when subject to external signal inputs.
Particular emphasis is put on the interplay between the internal properties of 
the open systems and the different possible inputs on them to deduce new 
functionalities of the modules.
We use discrete--time, piecewise--affine and piecewise--contracting models with 
interactions of a regulatory nature to perform our study.

\bs {\it Keywords}: 
Regulatory Dynamics on Networks, Open Systems, Motifs, Modularity.

\ms {\it MSC:} 37N25, 37L60, 05C69.

\end{abstract}



\section{Introduction}~\label{section-introduction}

\ms The structure of genetic regulatory networks can be abstracted by directed graphs, where the nodes represent genes and the arrows (oriented edges) stand for their interactions through transcription/translation products.
These interactions may be either activations or inhibitions.
Given the large number of components in most networks of biological interest, 
connected by positive and negative feedback loops, the comprehension of the dynamics of a system is often difficult if not impossible.
In this context, {\it mathematical modeling} eventually supported by {\it computer tools} can contribute to the analysis of a regulatory network by allowing the biologist to focus on a restricted number of plausible hypotheses, or to easily read some observed features of the system.
We refer to~\cite{Alonbook2007, HdJ&L05} and references therein for an account of the huge and still growing literature on this topic.

\bs It has been evidenced that the existence of network~\emph{motifs}, a set of recurring patterns
inside large biological regulatory networks, shall give new insight on the understanding of the 
performances of the global network.
We refer to the review~\cite{AlonNature2007} (see also~\cite{Alonbook2007}) for a clear and complete
description of such a point of view and references therein for recent work in different specific
biological contexts.
Since most of the time these \emph{modules} \cite{modularity} also have input and output interactions 
relating to the rest of the network, they may be considered as open dynamical systems, meaning 
distinguished parts of larger 'closed' systems, the remaining of the network being an 'environment'.

\ms In the classical literature of Dynamical Systems, ``open'' or ``forced'' dynamical systems are 
defined as those for which the dynamics depends on internal rules as well as on inputs from the 
environement, then producing some outputs back to the environement.
The notion of open systems is also the groundwork of System Theory and many results concerning the 
possibility to control linear and nonlinear systems have been obtained in the last 
decades~\cite{S98,I85}.
Here we are interested in a description of the dynamics of these systems in all the parameter space 
and subjected to any external signal.

\ms When comparing these modules with the standard corresponding regulatory networks we are faced to 
new questions to understand their dynamical behavior and, therefore, their possible new 
functionalities.
Here we study some of these questions by searching first for general properties of such systems and 
by trying to identify the fundamental mechanisms at work in the open systems we deal with.
Although our definition of module is not the most general possible, as it will appear next, it seems 
that it contains all the cases treated in the literature and, remarkably, it allows simpler proofs in 
many cases.

\ms These general properties must be complemented in each particular case with further analysis to 
derive additional specific properties of a given module and we perform this task for some examples.

\ms What we found in the general case may be summarized as follows: each elementary input selects a 
particular subset of the phase space and a particular dynamical rule among a (finite) set of 
possibilities allowed by the module; therefore any (finite or infinite) sequence of inputs drives the 
dynamics of the module along a pre--determined maze of possible paths.
Simple sequences of inputs give rise to simple dynamics of the module and therefore to simple 
responses (outputs), but, since they fix the system in a ``corner" of its phase space, the system may 
acquire new functionalities as we shall see in some examples.
Moreover, because such constraints in phase space may depend of the values of the parameters of the 
system and not only of its structure (the pattern of interactions) it turns out that the structure 
cannot determine the functionality in all cases ~\cite{Ingram2006}.
It will also be clear from what follows that, on the contrary, in some cases it can happen that the 
type of response is not essentially affected by a change of parameters, and only particular aspects 
do, {\it e.~g.} speeding or slowing of the response.
Interestingly, when a module can perform more than one function, each one is generally robustly 
provided.
This fact does not exclude the possibility of a fine tuning of the interaction between the input and 
the internal dynamics that may end in an interesting interplay between both.

\ms Despite the fact that the proofs of our results is done for a special type of model of regulatory 
networks, i. e. discrete time dynamical systems \cite{HdJ&L05,CFML06,LU06,VL05}, we believe that 
these results have a more general thrust, for the underlying mechanisms are common to other types of 
dynamical systems, as ODEs for instance.

\ms Besides, discrete--time models provide a simple framework where the consequences of interaction 
delays are already included and they largely benefit from tools and techniques in the Dynamical 
Systems theory \cite{HdJ&L05,CFML06}.

\ms We study two different types of modules.
The first one, called regulatory cascades (RC), is defined as any open network without (internal) 
circuits.
The second type, called forced circuits (FC), is any open circuit.
The general case can be build as a combination of such elementary pieces and at the end we sketch the 
analysis of one such mixed RC-FC module on the bases of our previous results.

\ms For the RC modules we give a complete description of the dynamics.
We specify, in particular, the output as a function of the input knowing the structure and the 
internal set of parameters of the module.
It turns out that the dynamics of such RC works as a special kind of cellular automata and, due to a 
celebrated result in \cite{Lindgren1990}, they can emulate a Universal Turing Machine!

\ms For the FC modules, we give a complete description of the dynamics for the cases of one and two 
units in the fundamental case of constant inputs and for all the parameter values of the circuits.
Using the exponential contracting rate of the system, the behavior for more general input signals 
follows.
Here the input level fixes the dynamics of the module in a subset of the phase space allowing 
dynamical regimes that do not correspond to the typical autonomous version.

\ms The relation between the transducer like point of view used in the RC analysis and the 
constrained phase space used for the FC is done by symbolic dynamics.

\ms It is worth to notice that, as we shall see, modules that we show to perform different dynamical 
behaviors may be functionals in different situations~\cite{Mangan&AlonPNAS2003}, therefore allowing 
to perform different functions either for given values of parameters through different inputs, or 
vice--versa.

\ms The paper is organized as follows: section 2 recalls the basic properties of the models, section 
3 describes the dynamics of the regulatory cascades (RC), section 4 describes the dynamics of the 
forced circuits (FC) and section 5 shows how, in simple cases, it is possible to treat a mixed type 
module.
Section 6 is concerned with final comments and outlooks.

\ms

\bs \section{Description of the Model and General Properties}~\label{section-defintions}\


\subsection{Discrete--time regulatory networks}\ 

\ms We consider a special class of models, i. e. discrete--time regulatory networks.
These are discrete--time dynamical systems on a network (see~\cite{HdJ&L05,VL05,LU06,CFML06}).
By a network we mean a digraph, with vertices in a given finite set $V$ and with arrows (oriented 
links) taken from another given finite set $A\subseteq V \times V$, together with a set of additional 
characteristics we introduce next.

\ms Vertices account for interacting units carrying a certain activity level (a scalar in $[0,1]$ 
associated to each unit $v\in V$), depicting the product of a gene, and the arrows account for the 
interactions between them.

\ms Although the results presented below hold for more general interactions (as for instance 
multiplicative interactions) we restrict ourselves to additive inhibitory--activating interactions 
for the sake of simplicity (see~\cite{VL05,LU06,CFML06}).

\ms The model is defined as follows:
to each arrow $(u,v)\in A$ we associate an interaction threshold $T_{uv}\in [0,1]$, a sign 
$\sigma_{uv}\in \{-1,1\}$ indicating whether the action of $u$ over $v$ is an activation 
$(\sigma_{uv}=1)$ or is an inhibition $(\sigma_{uv}=-1)$, and a coupling strength 
$\kappa_{uv}\in [0,1]$.
The activity level of the network at time $t\in\nn$ is specified by the collection of the 
$\x_v^t\in [0,1]$, expressing the activity of each unit $v\in V$ at time $t\in \nn$.

\ms Denoting $I(v):=\{u\in V:\ (u,v)\in A\}$ the set of vertices acting over a vertex $v\in V$, the 
activation at time $t+1$ of the unit $v\in V$ is given by
\begin{equation}~\label{evolution}
\x_v^{t+1}:=a\x_v^t+(1-a)\sum_{u\in I(v)}\kappa_{uv} 
H(\sigma_{uv}(\x_u^t-T_{uv})).
\end{equation}
The constant $a\in [0,1)$ appearing in the equation plays the role of a degradation rate.
In absence of interaction, the activity level of a unit $v$ decreases exponentially fast to 
zero, $\x_v^t=a^t\x_v^0$. In the interaction terms, $H$ represents the Heaviside (step) 
function, $H(x)=1$ for $x>0$ and $H(x)=0$ for $x\leq 0$.
Hence each interaction term is a piecewise constant function whose value changes whenever 
one of the coordinates $\x^t_u$ crosses its own threshold $T_{uv}$.

\ms Without loss of generality the coupling strengths are normalized by 
$\sum_{u\in I(v)}\kappa_{uv}=1$ for each $v\in V$.

\ms In~\cite{CFML06} we described the dynamics of some of this networks in great details.
We shall come back to these results when comparing with their open versions.

\bs

\subsection{Open networks}~\label{Open_networks} \

\ms As mentioned above an open network is defined as part of a larger regulatory network 
together with the corresponding incoming and outgoing arrows, respectively from and to the 
rest of the network.
\emph{Motifs} are such open networks that appear more often than the expected frequency in 
some null statistical graph model~\cite{Milo&al02, AlonNature2007}.
In the present paper we study open subnetworks without any reference to their possible 
abundance inside a larger regulatory network and we rather call ''module'' any ('simple') 
open subnetwork.

\ms Any collection of $N$ vertices $V_{\rm mod}\subsetneq V$ of a network $V$ defines an 
open network (as long as not all the arrows with head in $V_{\rm mod}$ have tail in $V_{\rm mod}$).

\ms Formally an {\it open regulatory network} or a {\it module} is a regulatory network with 
vertices $V_{\rm mod}\subsetneq V$, the internal units, and three kind of distinguished arrows:
\begin{itemize}
\item[{\it (1)}] the incoming arrows, denoted $A_{\rm in}$ is the set 
$\{(u,v) : \ u\in V \setminus V_{\rm mod}, \ v\in V_{\rm mod}\}$,
\item[{\it (2)}] the inner arrows denoted $A_{\rm mod}$ is the set 
$\{(u,v) : \ u\in V_{\rm mod}, \ v \in V_{\rm mod}\}$ and,
\item[{\it (3)}] the outgoing arrows denoted $A_{\rm out}$ the set 
$\{(u,v) : \ u\in V_{\rm mod}\ , \ v\in V\setminus V_{\rm mod}\}$.
\end{itemize}

\ms The dynamics of the module is defined using the same rule (see Eq.~\ref{evolution}) 
as for the general case. However, as we consider the module as an open system (as if there 
were no feedback loops from  $V_{\rm mod}$ to  $V_{\rm mod}$ through 
$\Vext:=V\setminus V_{\rm mod}$) we may replace the exact knowledge of $\x^{t}_u$ 
for $u\in\Vext$ by the one of the corresponding symbol: 
$\theta_{uv}^t:=H(\sigma_{uv}(\x_u^t-T_{uv}))$.

\ms Therefore, the state of the system at a given time $t\in \nn$ is determined by a vector 
$\x^t\in [0,1]^{\#V_{\rm mod}}$ and the imposed external activation levels $\theta_{uv}^t$ for 
all $(u,v)\in A_{\rm in}$. We refer to the array of external activations, $\{\theta_{uv}^t$ 
for all $(u,v)\in A_{\rm in}\}$, as the {\it input code $\theta_{\rm in}^t$}.

\ms In the case of an open system, the evolution rule~\eqref{evolution} reads:
\begin{equation}~\label{eqdyn}
\x^{t+1}_v:=a \x^{t}_v+(1-a)D_v\left(\x^t, \theta_{\rm in}^t\right)
\end{equation}
with
\begin{equation}~\label{D}
D_v\left(\x^t, \theta_{\rm in}^t\right):=\sum_{u\in I(v)\cap 
V_{\rm mod}}\kappa_{uv} H(\sigma_{uv}(\x_u^t-T_{uv})) + 
\sum_{u\in I(v)\cap\Vext}\kappa_{uv} \theta_{uv}^t
\end{equation}

\ms The state of the internal units determine the activation of the internal arrows $A_{\rm mod}$.

\ms We put together this data in the internal code
\begin{equation}~\label{internal-code}
\theta_{\rm mod}^t:=\left(H(\sigma_{uv}(\x^t_u-T_{uv})):\ (u,v)\in A_{\rm mod} \right).
\end{equation} 

\ms Finally, the influence of the open subnetwork $(V_{\rm mod},A_{\rm mod})$ over its environment is codified in the sequences of output activations, which we group in {\it the output code}
\begin{equation}~\label{output-code}
\theta_{\rm out}^t:=\left(
H(\sigma_{uv}(\x^t_u-T_{uv})):\ (u,v)\in A_{V_{\rm mod}\to\Vext} \right).
\end{equation}

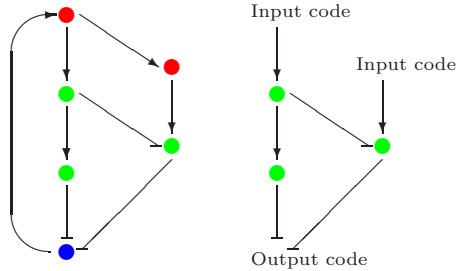
\begin{figure}[h]
\setlength{\unitlength}{0.7truecm}
\begin{center}
\begin{picture}(9,6)(0,0)
\put(2,5.25){\color[rgb]{1,0,0}\circle*{0.3}}
\put(2,3.75){\color[rgb]{0,1,0}\circle*{0.3}}
\put(2,2.25){\color[rgb]{0,1,0}\circle*{0.3}}
\put(2,0.75){\color[rgb]{0,0,1}\circle*{0.3}}

\put(2,5){\vector(0,-1){1}}
\put(2,3.5){\vector(0,-1){1}}
\put(2,3.5){\vector(0,-1){1}}
\put(2,2){\line(0,-1){1}}
\put(1.9,1){\line(1,0){0.2}}

\put(2.25,5.25){\vector(3,-2){1.5}}
\put(2.25,3.75){\line(3,-2){1.5}}
\put(4,2.5){\line(-1,-1){1.7}}
\put(2.2,0.8){\line(1,0){0.2}}

\put(4,4.25){\color[rgb]{1,0,0}\circle*{0.3}}
\put(4,2.75){\color[rgb]{0,1,0}\circle*{0.3}}
\put(4,4){\vector(0,-1){1}}
\put(3.6,2.75){\line(1,0){0.2}}

\put(6,3.75){\color[rgb]{0,1,0}\circle*{0.3}}
\put(6,2.25){\color[rgb]{0,1,0}\circle*{0.3}}

\put(6,5){\vector(0,-1){1}}
\put(6,3.5){\vector(0,-1){1}}
\put(6,2){\line(0,-1){1}}
\put(5.9,1){\line(1,0){0.2}}
\put(6.2,0.8){\line(1,0){0.2}}

\put(6.25,3.75){\line(3,-2){1.5}}
\put(8,2.5){\line(-1,-1){1.7}}

\put(5.5,5.2){\tiny Input code}
\put(7.5,4.2){\tiny Input code}
\put(8,2.75){\color[rgb]{0,1,0}\circle*{0.3}}
\put(8,4){\vector(0,-1){1}}
\put(7.6,2.75){\line(1,0){0.2}}
\put(5.5,0.5){\tiny Output code}

\put(1.7,3){\oval(1.5,4.5)[l]}
\put(1.6,5.25){\vector(1,0){0.2}}

\end{picture}
\ms
\caption{\small Open subnetwork (green) extracted from a larger one 
(red, green and blue). Hammer--like arrows represent inhibitory interactions, 
standard arrows activatory ones.}
\label{openexample}
\end{center}
\end{figure}

\ms In Figure \ref{openexample} we give an example of a module together with its external complement.

\ms 

\ms From Equations~\eqref{eqdyn} and (\ref{D}), we see that each possible input code 
$\theta:=\theta_{\rm in}$ uniquely determines an affine contraction 
$F_{\theta}:[0,1]^{\#V_{\rm mod}}\to [0,1]^{\#V_{\rm mod}}$ given by \begin{equation}~\label{Fteta}
F_{\theta}(\x)=a\x+(1-a)D_v\left(\x, \theta_{\rm in}\right).
\end{equation} 

\ms The collection of all these affine contractions defines an iterated function system (IFS)
\begin{equation}~\label{IFS}
\F:=\left\{F_{\theta}:[0,1]^{\#V_{\rm mod}}\to [0,1]^{\#V_{\rm mod}}:\ 
\theta\in \{0,1\}^{\#A_{\rm in}}\right\},
\end{equation}
from where the dynamics reads:
\begin{equation}~\label{IFSdyn}
\x^{t+1}=F_{\theta^{t-1}}\circ\cdots\circ F_{\theta^0}(\x^0),
\end{equation}
with attractor
\begin{equation}~\label{attractor}
\Omega_{\F}:=\bigcap_{t\geq 0} 
\bigcup_{\theta^0\cdots\theta^{t-1}\in\left(\{0,1\}^{\#A_{\rm in}}\right)^t}
F_{\theta^{t-1}}\circ\cdots\circ F_{\theta^0}([0,1]),
\end{equation}

\ms 
\begin{note}\label{long-remark} 
\begin{rm} The form in Equation~\eqref{D} is not the most general for which the results below 
can be proved.
Notice however that by a proper choice of the internal and external thresholds it already 
includes the cases of the AND and OR logical outputs~\cite{AlonNature2007}.
On the other hand, the essential ingredient needed for all the proofs is the collection 
of affine contractions indexed by the possible forcing codes, defining an iterated function 
system (IFS) as in~\eqref{IFS} and~\eqref{IFSdyn}.
It is also well known that such IFS are skew product dynamical systems~\cite{LARNOLD}.
However, because in our case we are interested in properties of very specific models, 
known results about the latter systems are only used to fix the general context of our 
study. They are, in particular, implicitly used when arguing on the genericity of some 
properties proved below.

\ms
In fact, for discrete--time regulatory networks, there is a way to understand the dynamics 
of an open system in terms of a collection of autonomous mirror systems.
For an open system $(V_{\rm mod},A_{\rm mod},A_{\rm in})$ and any constant forcing code: 
$\theta^0\cdots\theta^{t}\cdots$, with $\theta^{t}=\theta\, \forall t\in \nn$, let 
$\Omega_{\theta}$ be the corresponding attractor, named \emph{basic attractor} and 
defined by Equation~\eqref{attractor}.
Now, for each constant forcing code, the system will evolve inside an invariant subset of 
its phase space as an autonomous system, up to an affine change of variables.
This invariant subset attracts all the trajectories starting outside.
The lasts correspond, by the same change of variables, to those starting outside 
$[0,1]$ in the closed system and for which the behavior is known~\cite{CFML06}.

\ms On the other hand, the attractors for the closed systems, and therefore for each constant
forcing of the open system, are unions of the so called {\it global} orbits, those that can 
be extended backward in time up to $ - \infty$ inside $[0,1]^{\#V_{\it mod}}$ (see~\cite{CFML06}).
The same is then true for every basic attractor of the open system.
The generic case, in measure sense in parameter space, is such that each of the basic attractors 
is uniformly bounded away from the discontinuities (the internal thresholds).
In this case, any orbit with constant forcing will approach the attractor exponentially fast, 
with rate $\log a$. Therefore, for a general forcing the corresponding orbit will wander around 
the basic attractors and will closely approach one if the forcing input stays constant during 
a sufficiently long duration, and the dynamical behavior of the system is clear in this case.
Moreover it tells us that $a$ is the main parameter controlling how fast the system respond 
to a new input signal. This suggest that $a$ may be experimentally estimated when the 
corresponding proteins are not actively degraded, as it is the case for most proteins in 
growing bacterial cells~\cite{AlonNature2007}.

\ms For the remainder exceptional cases of the parameter values, the attractor will be 
arbitrarily close to the discontinuities.
In this situation an orbit approaching a basic attractor will sometimes be close but on 
the opposite side of the attractor. This will cause an accident in the internal code followed 
by an unforeseeable length of time before the orbit to approach again the basic attractor.

\ms The rigorous description of the dynamics of all the possible internal codes of the 
attractor in this case is still an open mathematical problem.
\end{rm}
\end{note}

\ms Recall that a path in a network $(V, A)$ is a sequence of vertices 
$(v_0,v_1,\ldots,v_{\ell-1})$, such that $(v_i,v_{i+1})\in A$ for $0\leq i \leq \ell-2$.
If in addition $(v_{\ell-1},v_0)\in A$, the we said that $(v_0,v_1,\ldots,v_{\ell-1})$ is a 
cycle. Circuits are cycles with no repeating vertices.

\ms


\bs \section{Regulatory cascades}~\label{section-cascade}\

\ms
\subsection{Description of the regulatory cascade}~\label{subsection-cascade}\

\ms Among the open structures encountered in the analysis of biological networks, and that 
could be associated to certain functions, are the feed--forward loops, the dense 
overlapping regulons, and the diamonds (see~\cite{AlonNature2007}).
These motifs have a two--common feature once regarded as digraphs: {\it (a)} they are 
connected and {\it (b)} they do not contain cycles.
As mentioned in the introduction, and as we will show below, open sub--systems on these 
kind of networks behave as finite state translators, from the input to the output code, 
with a certain delay which in principle may depend on the input structure.

\ms A {\it regulatory cascade (RC)} is an open subsystem $(V_{\rm mod},A_{\rm mod})$ defined 
in a regulatory network, such that it is connected and with no cycles. It is clear that such 
a network cannot be strongly connected. The open subnetwork in Figure \ref{openexample} is 
an example.

\begin{figure}[h]
\setlength{\unitlength}{0.6truecm}
\begin{center}
\begin{picture}(9,7.5)(0,0)
\put(0,7.2){$\Theta_{\rm in}\equiv$}

\put(3,7){\vector(2,-3){1.85}}
\put(4.2,7){\line(1,-3){0.9}}
\put(5,4.3){\line(1,0){0.2}}
\put(7.5,7){\vector(-2,-3){1.85}}

\put(2.8,7.2){$\theta_{u_1v}^t$}
\put(4.1,7.2){$\theta_{u_2v}^t$}
\put(6.0,7.2){$\cdots$}
\put(7.5,7.2){$\theta_{u_nv}^t$}

\put(5.3,3.9){\color[rgb]{0,1,0}\circle*{0.5}} 
\put(5.7,3.6){$v$}

\put(5,3.5){\vector(-2,-3){1.85}}
\put(5.2,3.5){\line(-1,-3){0.92}}
\put(4.17,0.73){\line(1,0){0.2}}
\put(5.35,3.5){\vector(0,-1){2.83}}
\put(5.5,3.5){\line(2,-3){1.83}}
\put(7.24,0.73){\line(1,0){0.2}}

\put(0,0.2){$\Theta_{\rm out}\equiv$}

\put(3,0.1){$\theta_{vw_1}^t$}
\put(4.2,0.1){$\theta_{vw_2}^t$}
\put(5.2,0.1){$\theta_{vw_3}^t$}
\put(6.1,0.1){$\cdots$}
\put(7.5,0.1){$\theta_{vw_m}^t$}

\end{picture}
\ms
\caption{\small The elementary transducer: an open subsystem consisting of a single vertex 
with several input and output arrows. Lines with hammer--like heads represent inhibitory 
interactions, arrows activatory ones.}~\label{fig2}
\end{center}
\end{figure}
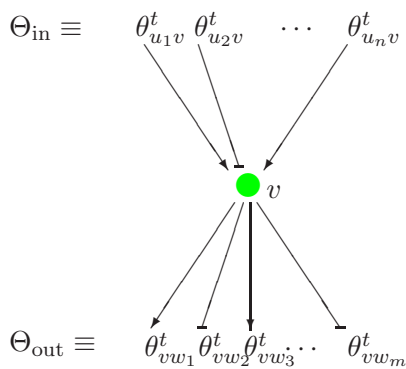

\ms The simplest RC (see Figure~\ref{fig2}) consists of a single internal vertex $v\in V$, 
forced by vertices $U:=\{u_1,\ldots,u_n\}\subset \Vext\equiv V\setminus\{v\}$, and affecting 
a collection of vertices $W:=\{w_1,\ldots,w_m\}\subset \Vext$.
Its functioning can be though as the translation of an input code 
$\theta_{\rm in}^t:= \left(H(\sigma_{uv}(\x^t_u-T_{uv})):\ u\in U \right)$, into an output 
code $\theta_{\rm out}^t:=\left(H(\sigma_{vw}(\x^t_v-T_{vw})):\ w\in W \right)$.
We assume that the vertex $v$ does not interact with itself, so that $(v_0,v_0)\notin A$.
We will refer to this open subsystem as as the {\it elementary transducer (ET)}.

\ms 

\ms Another elementary regulatory cascade is what we call a {\it regulatory chain RCh}.
It is defined over a linear network $\to v_1\to\cdots\to v_n\to $ consisting of a path 
connecting the input vertex ($v_1$) to the output one ($v_n$), with input and output 
arrows, $(u,v_1)$ and $(v_n,w)$ respectively.
Since to each interaction mode (activation/inhibition) it corresponds a sign (+1/-1), 
we can therefore associate the sign 
$\sigma:=\prod_{k=1}^{n-1}\sigma_{v_kv_{k+1}}\times\sigma_{v_n w}$ to the RCh 
$\to v_1\to\cdots\to v_n\to $.
As we will show below, the functioning of this chain as a transducer essentially depends 
of its sign.

\begin{figure}[h]
\setlength{\unitlength}{0.7truecm}
\begin{center}
\begin{picture}(4,5)(0,0)

\put(1.5,4.8){$\theta^t_{\rm in}$}
\put(3,4.8){\vector(0,-1){0.5}}

\put(3,4){\color[rgb]{0,1,0}\circle*{0.3}} 
\put(3.5,4){$u$}
\put(3,3.7){\vector(0,-1){1}}

\put(3,2.4){\color[rgb]{0,1,0}\circle*{0.3}}
\put(3.5,2.4){$v$}
\put(3,2.1){\vector(0,-1){1}}

\put(3,0.8){\color[rgb]{0,1,0}\circle*{0.3}}
\put(3.5,0.8){$w$}
\put(3,0.5){\vector(0,-1){0.5}}
\put(1.5,0){$\theta_{\rm out}^t$}

\put(2.7,2.35){\oval(1,3.3)[l]}
\put(2.7,0.6){\line(0,1){0.2}}

\end{picture}
\ms
\caption{\small Incoherent feedforward loop of the type 3}~\label{fig3}.
\end{center}
\end{figure}
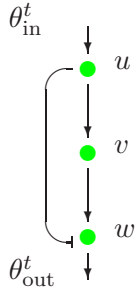

\ms A remarkable family of RCs are the so called {\it feedforward loops (FFL)}, 
consisting of three internal vertices connected as indicated in Figure~\ref{fig3}.
There are single input and output arrows, and two chains (elementary paths) connecting 
them. Taking this into account, the feedforward loops were classified as coherent and 
incoherent, depending on whether the sign of the chains composing them are of the same or 
opposite signs (see~\cite{AlonNature2007} for details).

\ms 

\ms As we will show below, it is convenient to decompose a given regulatory cascade into elementary transducers and regulatory chains.

\bs
\subsection{Dynamics of the regulatory cascade}~\label{subsection-dynamics-cascade}\

\ms We investigate how efficiently and robustly a regulatory cascade can work when 
transmitting information. Two mathematical questions are clearly related to this: what is 
the time length of the external signal needed by the cascade system to tie an output, and 
what is the possible dependence of the output on the internal state of the cascade at the 
time when the input is detected. In particular, our aim in this section is to determine the
conditions under which each output code of a RC depends only on the sequence of input codes 
received during a finite period of time, whose length depends on the structure and on the 
parameters of the RC. When this phenomenon occurs we say that the cascade resolves the 
output code with a finite delay. To this aim, we will first analyze in details the simplest 
RC, i. e. the elementary transducer. Since every RC can be decomposed into a collection 
of ET's, the conditions under which a general regulatory cascade resolves the code, and 
is functioning as a code transducer, will be deduced from the behavior of its elementary 
transducers.

\ms \subsubsection{\bf The elementary transducer}\label{paragraph-ET}\

\ms For the ET in Figure~\ref{fig2}, a temporal sequence of input codes 
$\Theta_{\rm in}:=(\theta_{\rm in}^t)_{t\in\nn}$, completely determines the evolution of 
the internal vertex:
\begin{equation}\label{elementary-transducer}
\x_{v}^{t+1}:=a\x_0^t+(1-a)\sum_{u\in U}\kappa_{uv}\theta_{uv}^t.
\end{equation}
Hence, the temporal sequence of output code 
$\Theta_{\rm out}:=(\theta_{\rm out}^t)_{t\in\nn}$ can be computed from this sequences of 
input codes and the initial condition $\x_v^{0}$.

\ms When the input signal has been present for an infinite time, the output code depends 
only on the infinite sequence of input signals. Otherwise, if the input signal started its 
action at time $t_0$, then the output code will also depend on $\x_v^{t_0}$, the activity 
level of the internal vertex at time $t_0$.

\ms According to Equation~\eqref{elementary-transducer}, from a sequence of input codes 
$\Theta_{\rm in}:=(\theta_{\rm in}^\tau)_{\tau=t_0}^{t}$, and an initial condition 
$\x_v^{t_0}$, we obtain
\begin{equation}\label{iterated-elementary-tranducer}
\x_v^t=a^{t-t_0}\x_v^{t_0}+(1-a)\sum_{\tau=t_0}^{t-1}a^{t-\tau-1}
                      \left(\sum_{u\in U}\kappa_{uv}\theta_{uv}^{\tau}\right),
\end{equation}
from which we compute the output code 
$\theta_{\rm out}^t:=
\left(H\left(\sigma_{vw}\left(\x_v^t-T_{vw}\right)\right):\ (v,w)\in A_{\rm out}\right)$.
Hence, in order to compute the temporal sequence of output codes 
$\Theta_{\rm out}:=(\theta_{\rm out}^t)_{t\in \nn}$, 
we have to determine the position of $\x_{v}^t$ with respect to the output thresholds 
$\{T_{vw}:\ (v,w)\in A_{\rm out}\}$, and so for each time $t\in\nn$.
To solve this problem we take into account the following.

\ms 
\paragraph{\bf \em 1.- The internal vertex IFS} Each possible input code 
$\theta_{\rm in}:=(\theta_{uv}:\ (u,v)\in A_{\rm in})$, uniquely determines an affine 
contraction $F_{\theta}:[0,1]\to [0,1]$ given by
\begin{equation}\label{theta-contraction}
F_{\theta}(\x)=a\x+(1-a)\sum_{(u,v)\in A_{\rm in}}\kappa_{uv}\theta_{uv}.
\end{equation}

\ms As already mentioned, the collection of all these affine contractions defines the 
iterated function system (IFS)
\begin{equation}\label{local-IFS}
\F:=\left\{F_{\theta}:[0,1]\to [0,1]:\ \theta\in \{0,1\}^{\#A_{\rm in}}\right\},
\end{equation}
with attractor
\begin{equation}\label{local-attractor}
\Omega_{\F}:=\bigcap_{t\geq 0}
\bigcup_{\theta^0\cdots\theta^{t-1}\in(\{0,1\}^{\#A_{\rm in}})^t}
F_{\theta^{t-1}}\circ\cdots\circ F_{\theta^0}([0,1]).
\end{equation}

\ms \paragraph{\bf \em 2.- Resolution of the output code} 
For each input sequence $(\theta_{\rm in}^\tau)_{\tau\geq 0}$ and all time 
$t\geq 1$, let 
$I_{\theta^0_{\rm in}\cdots\theta_{\rm in}^{t-1}} :=
F_{\theta_{\rm in}^{t-1}}\circ\cdots\circ F_{\theta_{\rm in}^0}([0,1])$.
The evolution of $\x_v^t$ is such that for each $t_0\in \nn$ fixed and all $t\geq t_0$ we 
have
\[\x_{v}^t\in
I_{\theta_{\rm in}^{0}\cdots\theta_{\rm in}^{t-1}}
\subset 
I_{\theta_{\rm in}^{t-t_0}\cdots\theta_{\rm in}^{t-1}}\equiv 
   [\x_v(\theta_{\rm in}^{t-t_0}\cdots\theta_{\rm in}^{t-1}),a^{t_0}
   +\x_v(\theta_{\rm in}^{t-t_0}\cdots\theta_{\rm in}^{t-1})],
\]
where 
\begin{equation}\label{x-of-theta}
\x_v(\theta_{\rm in}^{t-t_0}\cdots\theta_{\rm in}^{t-1})
:=(1-a)\sum_{\tau=t-t_0}^{t-1}a^{t-\tau-1}
      \left(\sum_{u\in U}\kappa_{uv}\theta_{uv}^{\tau}\right).
\end{equation}
the $t_0$ approximation to the activity level.
From Equation~\eqref{x-of-theta} it readily follows that 
\[
\left|\x_v^t-\x_v(\theta_{\rm in}^{t-t_0}\cdots\theta_{\rm in}^{t-1})\right|
\leq a^{t_0},
\]
for all $t > t_0$. Hence, for each $t_0$ fixed, the output code can be resolved for all the 
input codes $\theta_{\rm in}^{t-t_0}\cdots\theta_{\rm in}^{t-1}$ satisfying
\[
\min_{w\in W}\left| \x_v(\theta_{\rm in}^{t-t_0}\cdots\theta_{\rm in}^{t-1})
 -T_{uw}\right| > a^{t_0}.
\]
As mentioned above, in this case we say that the code is resolved with a finite delay $t_0$.
The set of input code sequences which can be resolved with a delay $t_0$ grows exponentially 
with $t_0$. Depending on the parameters of the system, the complement of this set could 
grow exponentially as well.

\ms There are two cases where we can show that the ET resolves the input code in a finite time.
The first case relies on an internal characteristic of the ET we call {\it internal separability}.
In this case, with probability $1$, the thresholds are separated from the attractor.
In the second case we assume that the input signal satisfies a property we name {\it low 
input complexity}. 

\ms \paragraph{\bf \em 3.- Internal separability} 
A possible simplification occurs when the attractor $\Omega_{\F}$ is a Cantor set.
This is the case if in Equation~\eqref{theta-contraction} the contraction rate $a$ is 
sufficiently small. In this cantorian case, for any $T\notin\Omega_{\F}$ there exists a 
{\it depth $t_T\in\nn$} such that 
$T\not\in \bigcup_{\theta^0\cdots\theta^{\tau-1}\in(\{0,1\}^{\#A_{\rm in}})^\tau} 
I_{\theta_{\rm in}^{0}\cdots\theta_{\rm in}^{\tau-1}}$, for each $\tau\geq t_T$.
Hence, if $\{T_{vw}:\ (v,w)\in A_{\rm out}\}\cap \Omega_\F=\emptyset$, which happens with 
probability 1, the maximal depth $t_0:=\max_{T}t_T$ is such that
$\theta_{\rm out}^t=
\left(
 H\left(\sigma_{vw}\left(
    \x_v(\theta_{\rm in}^{t-t_0}\cdots\theta_{\rm in}^{t-1})
                   \right)
   \right):\ (v,w)\in A_{\rm out}\right)$.
In this case the elementary transducer acts as a cellular automata, transforming sequences of 
input codes to sequences of output ones, with a delay $t_0$.
Indeed, we can define 
$\Phi:\left(\{0,1\}^{\#A_{\rm in}}\right)^{\nn}\to \left(\{0,1\}^{\#A_{\rm out}}\right)^{\nn}$, 
such that
\begin{equation}\label{simple-cellular-automata}
\Phi(\theta_{\rm in})_\tau:=
    \left(H\left(\sigma_{vw}\left(
        \x_v(\theta_{\rm in}^{t-t_0}\cdots\theta_{\rm in}^{t-1})
                   \right) \right):\ w\in W\right)
                            =\theta_{\rm out}^{\tau+t_0},
\end{equation}
with $\x_v(\theta_{\rm in}^{t-t_0}\cdots\theta_{\rm in}^{t-1})$ the $t_0$ approximation to 
the activity level defined in Equation~\eqref{x-of-theta}. Let us emphasize that in the 
cantorian case, the condition $\{T_{vw}:\ (v,w)\in A_{\rm out}\}\cap \Omega_\F = \emptyset$ 
holds with probability 1 with respect to Lebesgue, i.~e. a cantorian ET typically operates as 
a cellular automata. If so is the case, we say that the ET satisfies internal separability.
If on the contrary $\{T_{vw}:\ w\in W\}\cap \Omega_\F \neq \emptyset$, there would be for 
each $t\in \nn$ input codes $(\theta_{\rm in}^{0}\cdots\theta_{\rm in}^{t-1})$ that cannot 
be resolved in finite time.

\ms
\begin{example}\label{example1}
\begin{rm}
Let us illustrate the functioning of the elementary transducer in the internally separable 
case. Consider an ET with two input arrows $(u_1,v)$ and $(u_2,v)$, with coupling constants 
$\kappa_{u_1v}=1-\kappa_{u_2v}=2/3$, and a single output arrow $(v,w)$, as shown in 
Figure~\ref{fig-example1}. Let $T_{v,w}=83/150$ and $\sigma_{vw}=-1$.
Fix the contraction rate $a=1/5$. In this case the associated IFS is
\[
\F:=\{F_{00}(\x)=\x/5, F_{01}(\x)=\x/5+4/15, F_{10}(\x)=\x/5+8/15, F_{11}(\x)=\x/5+4/5\}.
\]
Its attractor $\Omega_{\F}$ is a Cantor set of box dimension 
$d_{\rm box}(\Omega_{\F})=\log(4)/\log(5)$.
In this particular case, the third approximant 
$\bigcup_{\theta_{\rm in}^0\theta_{\rm in}^1\theta_{\rm in}^{2} 
\in(\{0,1\}^{\#U})^3}I_{\theta_{\rm in}^{0}\theta_{\rm in}^1\theta_{\rm in}^{2}}$ 
of the IFS's attractor is the disjoint union of 64 closed intervals of length $1/125$, indexed 
by codes in $(\{0,1\}^2)^3$. None of these 64 intervals contain the output threshold, therefore 
we have a depth $t_0=3$ in this case.
The translating cellular automata $\Phi(\{0,1\}^2)^\nn\to \{0,1\}^\nn$ is defined in this 
case by the local function 
$\phi:(\{0,1\}^{2})^3\to\{0,1\}$ such that
\[
\phi(\theta_{\rm in}^0\theta_{\rm in}^1\theta_{\rm in}^2)=
\left\{\begin{array}{ll} 
1 \text{ if } \theta_{\rm in}^0\in\{00,01\} \text{ or } 
\theta_{\rm in}^0\theta_{\rm in}^1\theta_{\rm in}^2\in 
\{(10,00,00),(10,00,01)\},  \\
0 \text{ otherwise.}\end{array}\right.
\] 
Here, for instance, all input sequences in $\{11,01\}^\nn$  produce the same output sequence 
$000\cdots$.

\ms 
\begin{figure}[h]
\setlength{\unitlength}{0.7truecm}
\begin{center}
\begin{picture}(14,7)(0,0)
\put(0,5.8){$\Theta_{\rm in}\equiv$}
\put(1.5,5.4){\vector(1,-1){1.8}}
\put(1.4,5.8){$\theta_{u_1v}$}
\put(5.5,5.4){\vector(-1,-1){1.8}}
\put(5.3,5.8){$\theta_{u_2v}$}
\put(3.5,3.4){\color[rgb]{0,1,0}\circle*{0.3}}
\put(4,3.4){$v$}
\put(3.5,3.1){\line(0,-1){1.7}}
\put(3.4,1.4){\line(1,0){0.2}}
\put(3.9,1.3){$\Theta_{\rm out}$}

\linethickness{0.045mm}
\put(8,1){\vector(0,1){5.5}}
\put(8,1){\vector(1,0){5.5}}
\put(13,0.5){$x$}
\put(13,6){\line(-1,0){5}}
\put(13,6){\line(0,-1){5}}

\put(13.2,6){$F_{11}$}
\put(13.2,4.66){$F_{10}$}
\put(13.2,3.33){$F_{01}$}
\put(13.2,2){$F_{00}$}
\put(7,3.77){$T_{vw}$}

\put(8,1){\line(5,1){5}}
\put(8,2.33){\line(5,1){5}}
\put(8,3.66){\line(5,1){5}}
\put(8,5){\line(5,1){5}}

\put(8,1){\color[rgb]{0.8,0.8,0.8}\line(1,1){5}}

\put(8,3.77){\color[rgb]{1,0,0}\line(1,0){5}}
\end{picture}
\ms
\caption{\small The elementary transducer of Example~\ref{example1} and 
its associated IFS.}\label{fig-example1}
\end{center}
\end{figure}
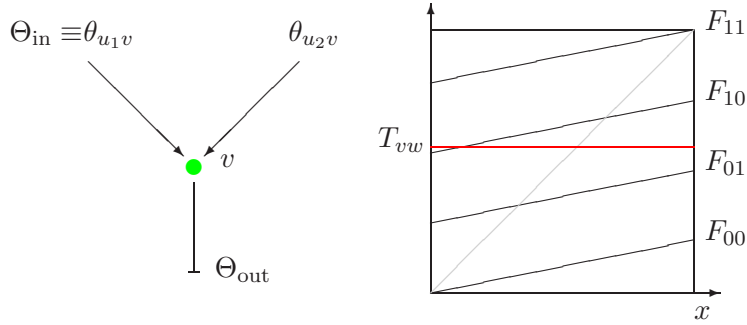
\ms
\end{rm}
\end{example}

\ms 
\paragraph{\bf \em 4.- Low input complexity} The attractor $\Omega_\F$ is a Cantor set if and 
only if $\cup_{\theta\in\{0,1\}^{\#A_{\rm in}}}F_\theta([0,1])\subsetneq [0,1]$.
When on the contrary $\bigcup_{\theta\in\{0,1\}^{\#A_{\rm in}}}F_\theta([0,1])$ covers $[0,1]$, 
even if the number of input codes that can be resolved with a delay $t_0$ grows exponentially 
fast with $t_0$, the cardinality of its complement could also grow exponentially fast.
In this case, in order to resolve the output code in finite time, we consider a particular 
class of input sequences.

\ms The {\it temporal complexity} of a sequence of input codes 
$\Theta_{\rm in}\equiv(\theta_{\rm in}^t)_{t\geq 0}$ is defined as
\begin{equation}\label{complexity}
C_{\Theta_{\rm in}}(t)\equiv
\#\left\{\theta_{\rm in}^{\tau}\cdots\theta_{\rm in}^{\tau+t}:\ 
\tau\in\nn\right\}.
\end{equation}
If the sequence of input codes $\Theta_{\rm in}$ is such that for a fixed $k\in \nn$ and 
all $t$ sufficiently 
large, $C_{\Theta_{\rm in}}(t)\leq t^k$ (in this case we say that the sequences of input codes 
have polynomial complexity), then the output code can be typically resolved with a finite delay.
We are interested in the polynomial case due to the fact that strongly contractive regulatory
networks produce sequences of codes with polynomial complexity~\cite{LU06}.
We conjecture that in all cases the sequences of codes produced by a regulatory network have 
polynomial complexity.

\ms We state our result in this case as follows. For sequences of input codes $\Theta_{\rm in}$ 
satisfying $C_{\Theta_{\rm in}}(t)\leq t^k$ for a fixed $k\in \nn$ and all $t$ sufficiently large, 
then, with probability 1 in the output thresholds, there exists a fixed delay time $t_1\geq t_0$ 
such that
\[
\theta_{\rm out}^{t}=
    \left(H\left(\sigma_{vw}\left(
        \x_v(\theta_{\rm in}^{t-t_1}\cdots\theta_{\rm in}^{t-1})
                   \right) \right):\ w\in W\right),
\]
with $\x_v(\theta_{\rm in}^{t-t_1}\cdots\theta_{\rm in}^{t-1})$ as defined by~\eqref{x-of-theta}.
The ET typically functions as a cellular automata when restricted to sequences of input codes with 
polynomial complexity.
In this case we say that the system has low input complexity.

\ms 
\begin{note}
\begin{rm}
The behavior of any totalistic cellular automaton can be obtained from an internally separable ET.
Since totalistic cellular automata can simulate any Turing machine~\cite{Gordon87}, it follows that 
ET's have universal computing capabilities.
\end{rm}
\end{note}

\bs
\subsubsection{\bf The Regulatory Chain (RCh)}\

\ms As mentioned above, a regulatory chain is an open regulatory network defined over a linear 
digraph 
$\to v_1\to\cdots\to v_n\to $, connecting the input vertex ($v_1$) to the output one ($v_n$), with 
input and output arrows $(u,v_1)$ and $(v_n,w)$ respectively.
To this chain we associate the sign 
$\sigma:=\prod_{k=1}^{n-1}\sigma_{v_kv_{k+1}}\times\sigma_{v_nw}$.
The functioning of this chain as a transducer essentially depends of this sign.

\ms We have two possibilities depending on the common contraction rate $a\in [0,1)$.
The simplest one occurs when $a < 1/2$, in which case all the vertices $v_1,\ldots,v_n$ 
considered as elementary transducers typically act as a cellular automata.
Indeed, if $a < 1/2$ to each vertex we associate the same dyadic IFS
\begin{equation}
\F_{\rm dyadic}:=\left\{F_{\theta}:[0,1]\to [0,1]:\ \theta\in \{0,1\}\right\},
\end{equation}
with attractor
\begin{equation}\label{local-attractor}
\Omega_{\rm dyadic}:=\bigcap_{t\geq 0}
\bigcup_{\theta^0\cdots\theta^{t-1}\in \{0,1\}^t}
F_{\theta^{t-1}}\circ\cdots\circ F_{\theta^0}([0,1]).
\end{equation}
This is a Cantor set with box dimension $\log(2)/\log(a^{-1})$.

\ms In the typical case, when 
$\{T_{v_kv_{k+1}}:\ 1\leq k\leq n+1\}\cap \Omega_{\rm dyadic}=\emptyset$,
we can associate, as before, to each internal vertex a depth 
\begin{equation}\label{chain-depth}
t_k:=\max\left\{t\geq 1: T_{v_kv_k+1}\in 
\cup_{\theta^0\cdots\theta^{t-1}\in \{0,1\}^t}
F_{\theta^{t-1}}\circ\cdots\circ F_{\theta^0}([0,1])
\right\}.
\end{equation}
In the present case we can recursively define the $t_k$ approximation to the activity level, 
and the internal code as follows:
\begin{eqnarray}\label{recursive-x-of-theta}
\x_{v_1}(\theta_{\rm in}^{t-t_1}\cdots\theta_{\rm in}^{t-1})
&:=&(1-a)\sum_{\tau=t-t_1}^{t-1}a^{t-\tau-1}\theta_{\rm in}^{\tau}\\
\theta_{v_kv_{k+1}}&:=&
H\left(\sigma_{v_kv_{k+1}}\left(\x_{v_k}
  (\theta_{v_{k-1}v_k}^{t-t_k}\cdots\theta_{v_{k-1}v_k}^{t-1})-T_{v_{k-1}v_k}
  \right)\right) \nonumber \\ 
\x_{v_{k+1}}(\theta_{v_kv_{k+1}}^{t-t_{k+1}}\cdots\theta_{v_kv_{k+1}}^{t-1})
&:=&(1-a)\sum_{\tau=t-t_{k+1}}^{t-1}a^{t-\tau-1}\theta_{v_kv_{k+1}}^{\tau},
\nonumber
\end{eqnarray}
for $1\leq k < n$.
According to this, the $(k+1)$--th internal vertex, considered as an ET, works as  the cellular automata $\Phi_k:\{0,1\}^\nn\to\{0,1\}^\nn$, such that
\begin{equation}\label{cellular-chain}
\Phi_k(\Theta)_{t}=\left\{\begin{array}{ll}
(1-\sigma_{v_{k+1}v_{k+2}})/2 & \text{ if }
\x_{v_{k+1}}(\theta^{t-t_{k+1}}\cdots\theta^{t-1}) 
                                < T_{v_{k+1}v_{k+2}}\\
(1+\sigma_{v_{k+1}v_{k+2}})/2 & \text{ otherwise.}\end{array}\right.                     
\end{equation} 
It follows that the output code of the whole RCh can be resolved with a delay 
$t_0:=\sum_{k=1}^nt_k$, by using the composition
$\Phi:=\Phi_n\circ\Phi_{n-1}\circ\cdots\circ\Phi_1:\{0,1\}^\nn\to\{0,1\}^\nn$.

\ms The structure of each one of the cellular automata $\Phi_k:\{0,1\}^\nn\to\{0,1\}^\nn$ 
is such that if
$\theta_{\rm in}^{t-t_0}\cdots\theta_{\rm in}^{t-1}$ is a constant sequence, then
\begin{equation}\label{sign-cellular-chain}
\Phi(\Theta)_t=\left\{\begin{array}{ll}
(1-\sigma)/2 & \text{ if } 
       \theta_{\rm in}^{t-t_0}\cdots\theta_{\rm in}^{t-1}=00\cdots 0,\\
(1+\sigma)/2 & \text{ if } 
       \theta_{\rm in}^{t-t_0}\cdots\theta_{\rm in}^{t-1}=11\cdots 1,\end{array}
\right.
\end{equation}
where $\sigma=\prod_{k=1}^{n-1}\sigma_{v_kv_{k+1}}\times\sigma_{v_nw}$ is the sign 
of the chain as defined above.

\ms As for the ET, even if the dyadic attractor $\Omega_{\rm dyadic}$ is not a 
Cantor set, the output code of the regulatory chain can be typically resolved 
for sequences of 
low input complexity.
In this case the action can also be obtained as the composition 
$\Phi:=\Phi_n\circ\Phi_{n-1}\circ\cdots\circ\Phi_1:\{0,1\}^\nn\to\{0,1\}^\nn$.

\ms \subsubsection{\bf The general regulatory cascade}~\label{The general regulatory cascade}\

\ms As mentioned above a regulatory cascade is defined over a connected digraph 
with no cycles. The vertices of this digraph can be hierarchically organized, so 
that the input code for vertices in the $k$--th level are the output code of the 
vertices in the $(k-1)$--th level.
This way, following the same idea as in the analysis of the regulatory chain, 
we can resolve the output code of the cascade by using a composition of cellular 
automata associated to the vertices of the cascade.


\ms The hierarchy of vertices is the following.
On top we have the root vertices, 
\begin{equation}\label{root-vertices}
V_{\rm root}:=\{v\in V_{\rm mod}:\ (u,v)\in A \Rightarrow u\in V_{\rm ext}\}, 
\end{equation}
which are the vertices in $V_{\rm mod}$ all of whose incoming arrows have 
tail in the set of external vertices.
Let $V_0=V_{\rm root}$, $U_0=V_0$, and for each $k\geq 1$ we define
\begin{eqnarray}\label{level-k-vertices}
V_k&:=&\{v\in V_{\rm mod}:\ (u,v)\in A \Rightarrow u\in U_{k-1}\cup \Vext\},\\
U_k&:=&U_{k-1}\cup V_k.\nonumber
\end{eqnarray}
We can see $U_{k-1}$ as the union of all vertices up to the $(k-1)$--th level.
Then the $k$--th level, $V_k$, is composed by those vertices in $V_{\rm mod}$ all 
of whose incoming arrows have tail in levels lower than the $k$--th or in $\Vext$.
Since the underlying digraph has no cycles or loops, then these levels are nonempty.
Also, since the digraph $V_{\rm mod}$ is a finite set, there is a finite number 
of levels, all of them of finite size. Vertices of the last of theses levels are 
called \emph{leaf vertices}.
This last level can also be defined by
\begin{equation}\label{leaf-vertices}
V_{\rm leaf}:=\{v\in V_{\rm mod}:\ (v,w)\in A\Rightarrow w\in V_{\rm ext}\}.
\end{equation}
The depth $d$ of the regulatory cascade is the number of steps needed, starting 
from the root, to determine the leaf vertices, {\it i.~e.} $V_{d}=V_{\rm leaf}$.

\ms Because of this hierarchical structure, and taking into account 
Equation~\eqref{eqdyn}, the activity level $\x_v^t$, for $v\in V_k$, is given by
\begin{equation}\label{evolution-cascade}
\x_v^{t+1}:=a\x_v^t+(1-a)\left\{\sum_{u\in I(v)\cap U_{k-1}}\kappa_{uv} 
   H(\sigma_{uv}(\x_u^t-T_{uv})) +\sum_{u\in I(v)\cap\Vext }\kappa_{uv}\theta_{uv}^t\right\}.
\end{equation}
If, on the other hand, we can resolve the internal codes 
$(\theta_{uv}^t:\ u\in I(v)\cap U_{k-1})$, then the previous equation reduces to
\[
\x_v^{t+1}:=a\x_v^t+(1-a)\left\{\sum_{u\in I(v)\cap U_{k-1}}\kappa_{uv} \theta^t_{uv}
+\sum_{u\in I(v)\cap\Vext }\kappa_{uv}\theta_{uv}^t\right\}.
\]
Hence, all the internal vertices can be considered as ETs, and all what we discussed 
in paragraph~\ref{paragraph-ET} applies.
Once again we have the alternative between the internally separable and the 
non--separable case.

\ms \paragraph{\bf \em 1.- Internal separability for RC}
Once again, the simplest case occurs when the IFS associated to each one of the 
internal vertices,
\[
\F_v:=\left\{F_\theta(\x)=a\x+(1-a)\sum_{u\in I(v)}\kappa_{uv} \theta_{uv}:\
\theta \in \{0,1\}^{\#I(v)}\right\}
\]
has a Cantor set $\Omega_v$ as attractor.
In this simple case, with probability 1 none of the thresholds associated to internal 
arrows will lie inside its 
corresponding Cantor set, so that to the sequence of output codes from vertex $v$, 
$(\theta_{vv'}^t:\ (v,v')\in A)$ can be resolved with a finite delay $t_v$, by using 
a cellular automata $\Phi_v:(\{0,1\}^{\#I(v)})^\nn\to(\{0,1\}^{\#O(v)})^\nn$.
Here $I(v)$ and $O(v)$ are respectively the input and output set of the vertex $v$, 
and the cellular
automata $\Phi_v$ is defined in the same way as in Equation~\eqref{simple-cellular-automata}.
The input sequence of the cellular automata associated to the internal vertex $v\in V_k$ is 
obtained from the input sequence of codes via the action of the cellular automata $\Phi_u$ 
associated to vertices in levels lower than $k$.

\ms
\begin{example}\label{example} 
\begin{rm}
Let us illustrate how operates a regulatory cascade in the internally separable case 
by considering the incoherent feedforward loop of Figure~\ref{fig3}.
This RC contains a positive regulatory chain $\to u\to v\to$, and an negative arrow 
from $u$ to $w$.
Let us fix $a=1/5$, output threshold $T_{wv_{\rm out}}=1/2$, output sign 
$\sigma_{wv_{\rm out}}=1$, and coupling constants $\kappa_{uw}=1-\kappa_{vw}=2/3$.
In this way, the functioning of the vertex $w$ considered as an ET could be deduced 
from that of Example~\ref{example1}.

\ms Let $T_{uv}=22/25$, $T_{uw}=22/125$, and $T_{vw}=1/2$. 
The IFS associated to these vertices is
\[\F_u=\F_v=\{F_0x=x/5,F_1x=x/5+4/5\},\]
which has a dyadic Cantor attractor $\Omega_v$ with box dimension $\log(2)/\log(5)$.
The delays determined from the position of the thresholds with respect to the attractor 
are $t_u=2$ and $t_v=1$. 

\ms The internal code $(\theta_{uv}^t, \theta_{uw}^t)$ depends only on the input 
sequence $\Theta_{\rm in}$ and it is given by
\begin{eqnarray*}
\theta_{uv}^t&=&\left\{\begin{array}{ll}
                      0 & \text{ if } \theta_{\rm in}^{t-2}\theta_{\rm in}^{t-1},
                      \in\{00,01,10\}, \\
                      1 & \text{ if } \theta_{\rm in}^{t-2}\theta_{\rm in}^{t-1}=11.
                      \end{array}\right.\\
\theta_{uw}^t&=&\left\{\begin{array}{ll}
                      1 & \text{ if } 
                      \theta_{\rm in}^{t-3}\theta_{\rm in}^{t-2}\theta_{\rm in}^{t-1}
                      \in\{000,001,010\}, \\
                      0 & \text{ otherwise.}\end{array}\right.\\    
\end{eqnarray*}
For the RCh $\to u\to v\to$, the composition of $\Phi_v\circ\Phi_u$ allows to 
determine the internal code $\theta_{vw}^t$ directly from the input code $\Theta_{\rm in}$.
We obtain the following:
\[
\theta_{vw}^t=\left\{\begin{array}{ll}
                      0 & \text{ if } \theta_{\rm in}^{t-3}\theta_{\rm in}^{t-2}
                      \in\{00,01,10\}, \\
                      1 & \text{ otherwise.}\end{array}\right.
\]

\ms Finally, considering the vertex $w$ as an ET with two inputs we obtain an output code
\[
\theta_{\rm out}^t= \left\{\begin{array}{ll}  1 \text{ if } 
                                              \theta_{\rm in}^{t-4}\theta_{\rm in}^{t-3}
                                                                     \in\{00,01,10\},  \\
                                               0 \text{ if } 
                                           \theta_{\rm in}^{t-4}\theta_{\rm in}^{t-3}=11.
                       \end{array}\right.
\]
This feedforward loop operates as a cellular automata.
The output code can be resolved with a delay $t_w=4$.

\bs

\begin{center}
\begin{figure}
\begin{picture}(150,250)
\includegraphics[scale=0.8]{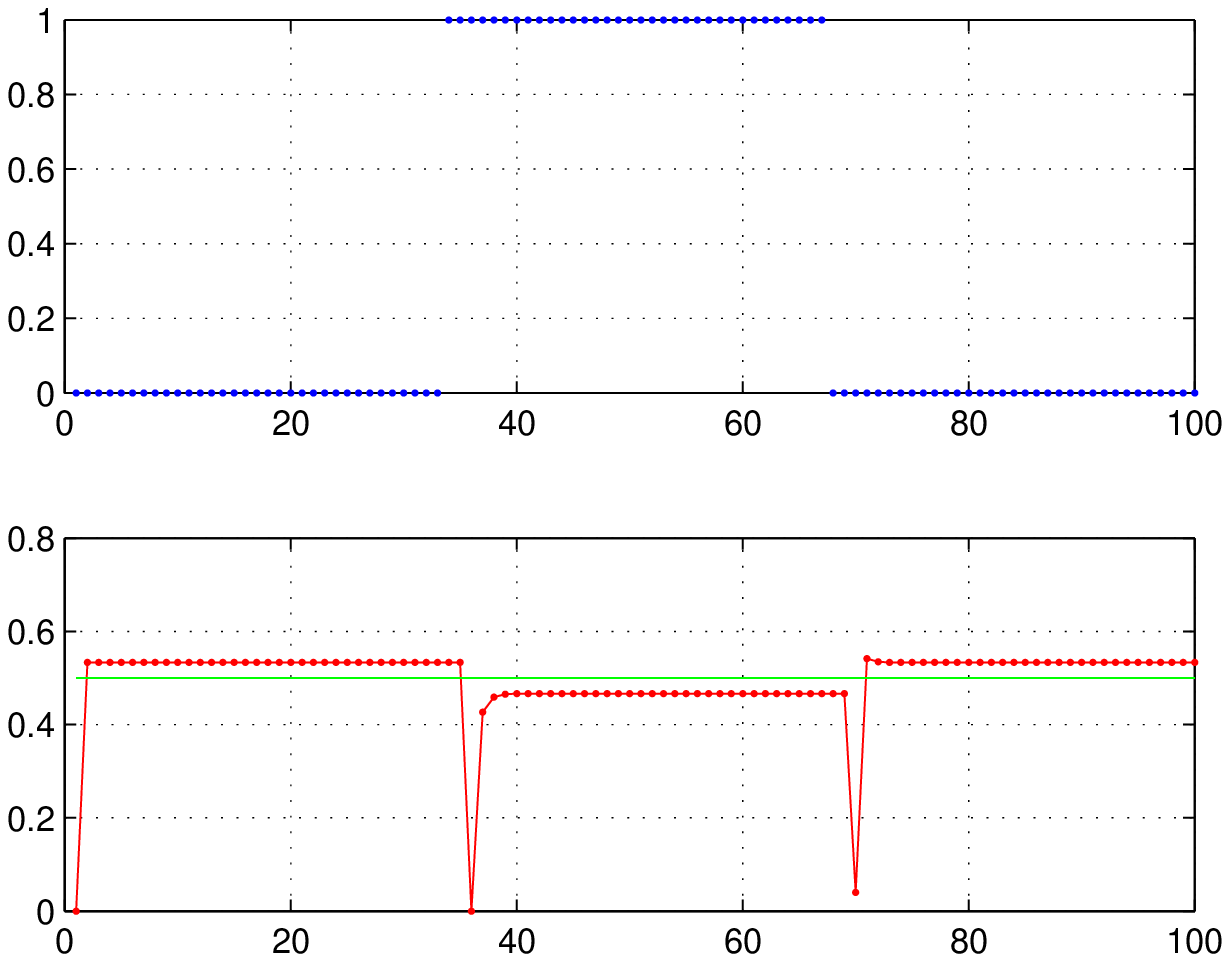}
\end{picture}
\begin{picture}(150,250)
\put(-130,80){\makebox(10,0){$T_{uw}$}}
\put(120,105){\makebox(10,0){$\x_w$}}
\put(150,5){\makebox(10,0){$t$}}

\put(120,210){\makebox(10,0){$\Theta_{\rm in}$}}
\put(150,125){\makebox(10,0){$t$}}
\ms
\end{picture}

\caption{\small Behavior of the incoherent feedforward loop. 
On top, the input signal $\theta_u^t$. Below, the state of the output vertex $\x^t_w$ 
responsible of the output signal. The output signal changes each time $\x_w^t$ crosses 
the output threshold indicated by a horizontal line in the figure.}~\label{timecascade}
\end{figure}
\end{center}

\end{rm}
\end{example}

\ms
\paragraph{\bf\em 2.- The general case}
The output code cannot be determined solely from the input code when at least one of 
the attractors $\Omega_v$ is not a Cantor set, or when one of the thresholds $T_{vv'}$ 
is contained in its corresponding attractor $\Omega_{v}$.
In those cases, for each depth $\tau\in\nn$ there are sequences of input codes such 
that the output code cannot be resolved in finite time.
The comment made in Paragraph~\ref{paragraph-ET} concerning sequences of input codes 
with low temporal complexity applies once again. In that case we can typically resolve 
the output code in finite time.

\ms \begin{note}[Comments on decrease of complexity]
\begin{rm} Each time we can resolve the output code in finite time, whether we 
are in the internally separable case or because the low input complexity holds, 
the RC operates as a cellular automata 
$\Phi:(\{0,1\}^{\#A_{\rm in}})^\nn\to(\{0,1\}^{\#A_{\rm out}})^\nn$ 
such that
\[
\Phi(\Theta_{\rm in})_t=\phi(\theta_{\rm in}^{t-t_0} 
\theta_{\rm in}^{t-t_0+1}\cdots\theta_{\rm in}^{t-1}).
\]
The symbolic complexity of the output code is the counting function 
$C_{\Theta_{\rm out}}:\nn\to\nn$ such that
\begin{equation}\label{out-temporal-complexity}
C_{\Theta_{\rm out}}(t):= \#\left\{\theta_{\rm out}^{\tau}\cdots\theta_{\rm in}^{\tau+t}:\ 
\tau\in\nn\right\},
\end{equation}
{\it i.~e.}, for each $t\in\nn$, $C_{\Theta_{\rm out}}(t)$ counts all different sequences 
of output codes of length $t$. In the present case, since the output code can be 
resolved with a delay $t_0$, we clearly have 
\begin{equation}\label{complexity-inequality}
C_{\Theta_{\rm out}}(t+t_0)\leq C_{\Theta_{\rm in}}(t)+\left(2^{\#W}\right)^{t_0}.
\end{equation} 
In the case of an exponentially increasing complexity the previous inequality 
ensures the non--increase of the entropy~\cite{LU06}.
\end{rm}
\end{note}

\ms
\bs \section{Forced Circuits}~\label{section-fcircuits} \

\ms Forced circuits (FC) are modules in a regulatory network such that the 
underlying graph  $(V_{\rm mod},A_{\rm mod})$ is a circuit.

\ms Examples of forced circuits are the (open) Negative Auto Regulator (NAR), 
the Positive Auto Regulator (PAR) and the network motifs with 
double--positive (or double--negative)--feedback loop~\cite{AlonNature2007}.

\ms In this section we focus on special cases of FC in order to show the 
strategy of analysis as well as the reach possibilities of dynamical behaviors 
displayed by such modules. Not surprisingly, each one may show different behaviors 
according to the inputs.
In particular, in most cases the dynamical behavior of the module (open system) is 
different from that of the same system when isolated.

\ms \subsection{The self--regulations}\

\ms Self-regulation occurs when a transcription factor acts as an inhibitor 
(self--inhibitor or NAR) or an enhancer (self-activation or PAR) of the transcription 
of its own gene. Self-regulation is a very common situation, for instance it is involved 
in over $40\%$ of known {\it E. coli} transcription factors~\cite{THPRCV00}.

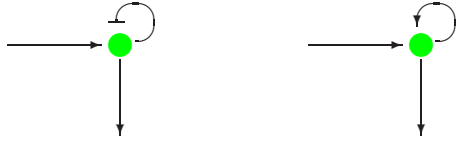
\begin{figure}[h]
\setlength{\unitlength}{1truecm}
\begin{center}
\begin{picture}(8,2.5)(0,0)
\put(1.5,1.5){\color[rgb]{0,1,0}\circle*{0.3}}
\put(0,1.5){\vector(1,0){1.25}}
\put(1.7,1.8){\oval(0.5,0.5)[r]}
\put(1.7,1.8){\oval(0.5,0.5)[tl]}
\put(1.35,1.8){\line(1,0){0.2}}
\put(1.5,1.3){\vector(0,-1){1}}

\put(5.5,1.5){\color[rgb]{0,1,0}\circle*{0.3}}
\put(4,1.5){\vector(1,0){1.25}}
\put(5.7,1.8){\oval(0.5,0.5)[r]}
\put(5.7,1.8){\oval(0.5,0.5)[tl]}
\put(5.45,1.85){\vector(0,-1){0.1}}
\put(5.5,1.3){\vector(0,-1){1}}

\end{picture}
\ms
\caption{\small  The open self--regulation, left: $\sigma_{vv} = -1$ for a self-inhibition; 
right: $\sigma_{vv} = +1$ for a self-activation.}
\label{supplementary_Cuer_1}
\end{center}
\end{figure}

\ms For a self--regulator subjected to a single input arrow as indicated in 
Figure~\ref{supplementary_Cuer_1}, thanks to the normalization condition 
$\kappa_{vv} + \kappa_{uv} = 1$, the Equations~\eqref{evolution} and~\eqref{eqdyn} read:
\begin{equation}~\label{dynSelfReg}
\x^{t+1}_v = a \x^t_v + (1-a) \left[ H \left( \sigma_{vv} (\x^t_v-T_{vv})
\right) + \kappa_{uv} \left( \theta^t_{uv} - H \left( \sigma_{vv} (\x^t_v-T_{vv})
\right) \right) \right],
\end{equation}
From~\eqref{evolution}, we see that for $\kappa_{uv} = 0$ one recovers the isolated 
self--regulation.

\ms Otherwise, for each given $\sigma_{vv}$  there are two extra parameters, the input 
intensity $\kappa_{uv}$ and the input signal sequence 
$\Theta_{\rm in}:=(\theta_{uv}^t)_{t \in \mathbb{N}}$
(also called ``exogene variable'').
As usual we write $H \left( \sigma_{vv} (\x_{v}-T_{vv}) \right) = \theta_{vv}$ and then 
we merge internal and external codes in a unique symbol: 
$\theta^t = \theta_{vv}^t\theta_{uv}^t$ ({\it internal on the left, 
input on the right}).

\ms As a first approximation, an easy way to visualize the different dynamics when 
changing the parameters and/or the input signal sequence is to show a diagram 
with the possible transitions among the $\theta$s at each time step 
(see Figure~\ref{OPEN_NEG_1D} (B)), known as dynamical 
graphs~\cite{Thomas1978,Edwards2001,DJong2002}.
In particular, it allows to locate forbidden paths in phase space and therefore localize 
the most robust bifurcations. Because residence times in each loop are not specified 
in this diagram, dynamical graphs carry only part of the dynamical information.

\ms We shall describe in the following the case of the self-inhibition in detail and 
then, for conciseness, we only sketch the case of the self-activation.

\ms 
\subsubsection{\bf The open self--inhibition}~\label{The open self-inhibition}\

\ms The dynamics of an isolated self-inhibitor, NAR, consists only of 
oscillations~\cite{CFML06}. In fact, whenever $0<T_{vv}<1$, this system is conjugated 
to a rotation on a circle with a rotation number $\nu(a,T)$ depending on the 
parameters~\cite{thesecoutinho,CFML06}.

\begin{figure}[h]
\begin{center}
\includegraphics[scale=0.3]{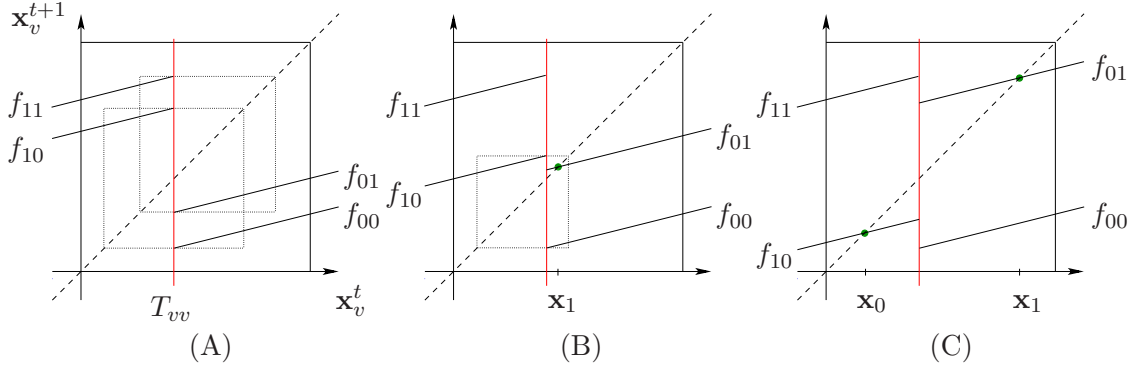}
\begin{picture}(400,0)
\put(113,13){\makebox(10,0){$\x^{t}_{v}$}}
\put(-5,115){\makebox(10,10){$\x^{t+1}_{v}$}}
\put(117,55){\makebox(10,10){$f_{01}$}}
\put(117,40){\makebox(10,10){$f_{00}$}}
\put(-10,65){\makebox(10,10){$f_{10}$}}
\put(-10,82){\makebox(10,10){$f_{11}$}}

\put(133,48){\makebox(10,10){$f_{10}$}}
\put(133,82){\makebox(10,10){$f_{11}$}}
\put(258,70){\makebox(10,10){$f_{01}$}}
\put(258,40){\makebox(10,10){$f_{00}$}}

\put(273,25){\makebox(10,10){$f_{10}$}}
\put(273,82){\makebox(10,10){$f_{11}$}}
\put(400,96){\makebox(10,10){$f_{01}$}}
\put(400,40){\makebox(10,10){$f_{00}$}}

\put(45,10){\makebox(10,0){$T_{vv}$}}
\put(193,12){\makebox(10,0){$\x_{1}$}}
\put(310,12){\makebox(10,0){$\x_{0}$}}
\put(369,12){\makebox(10,0){$\x_{1}$}}
\put(60,-5){\makebox(10,0){(A)}}
\put(200,-5){\makebox(10,0){(B)}}
\put(340,-5){\makebox(10,0){(C)}}
\end{picture}
\end{center}

\caption{\small {\bf The open self--inhibition}: graphs of the IFS
$F_{ \theta_{uv} } (\x_{v}) = a \x_{v} +(1-a) 
\left[H\left(T_{vv}-\x_{v}\right) + 
\kappa_{uv} \left(\theta_{uv}-H\left(T_{vv}-\x_{v}\right)\right) 
\right]
$
For, (A) Region I,  small input intensity $\kappa_{uv}$, (B) Region II, intermediate 
input intensity and (C) Region III, high input intensity. Possible fixed points are 
given by the intersection of the graph with the diagonal. 
The notation $f_{ij}, i,j=0,1$ stands for the branch of $F_{i}$ when 
$H \left(T_{vv}-\x_{v}\right)=i$ and $\theta_{uv}=j$.~\label{Figure_Cuer_2}
}
\end{figure}

\ms Figure~\ref{Figure_Cuer_2} illustrates the three possible cases of the dynamics 
for the open self-inhibitor, shown Figure~\ref{supplementary_Cuer_1}, when varying the 
input intensity. Accordingly, the parameter subspace 
$\{ (T_{vv}, a, \kappa_{uv}) \; : \; T_{vv} \in (0,1), a \in [0,1) \; \text{and} \; 
\kappa_{uv} \in [0,1] \}$ 
can be divided into three input intensity regions, corresponding to different 
dynamical  behaviors.
{\bf \em Region I:} if $\kappa_{uv} < T_{vv}$, {\bf \em Region II:} 
if $T_{vv}<\kappa_{uv} < 1-T_{vv}$ and {\bf \em Region III:} if $\kappa_{uv} > T_{vv}$.

\begin{note}  
\begin{rm}
Up to a change of $(\theta_{vv},\theta_{uv})$ in  $(1-\theta_{vv},1-\theta_{uv})$ in 
the dynamical graphs, it is enough to consider the cases where $T_{vv} <1-T_{vv}$.
\end{rm}
\end{note}

\ms 

\begin{figure}[h]
\begin{center}

\begin{picture}(400,200)
\put(50,-40){\includegraphics[scale=0.5]{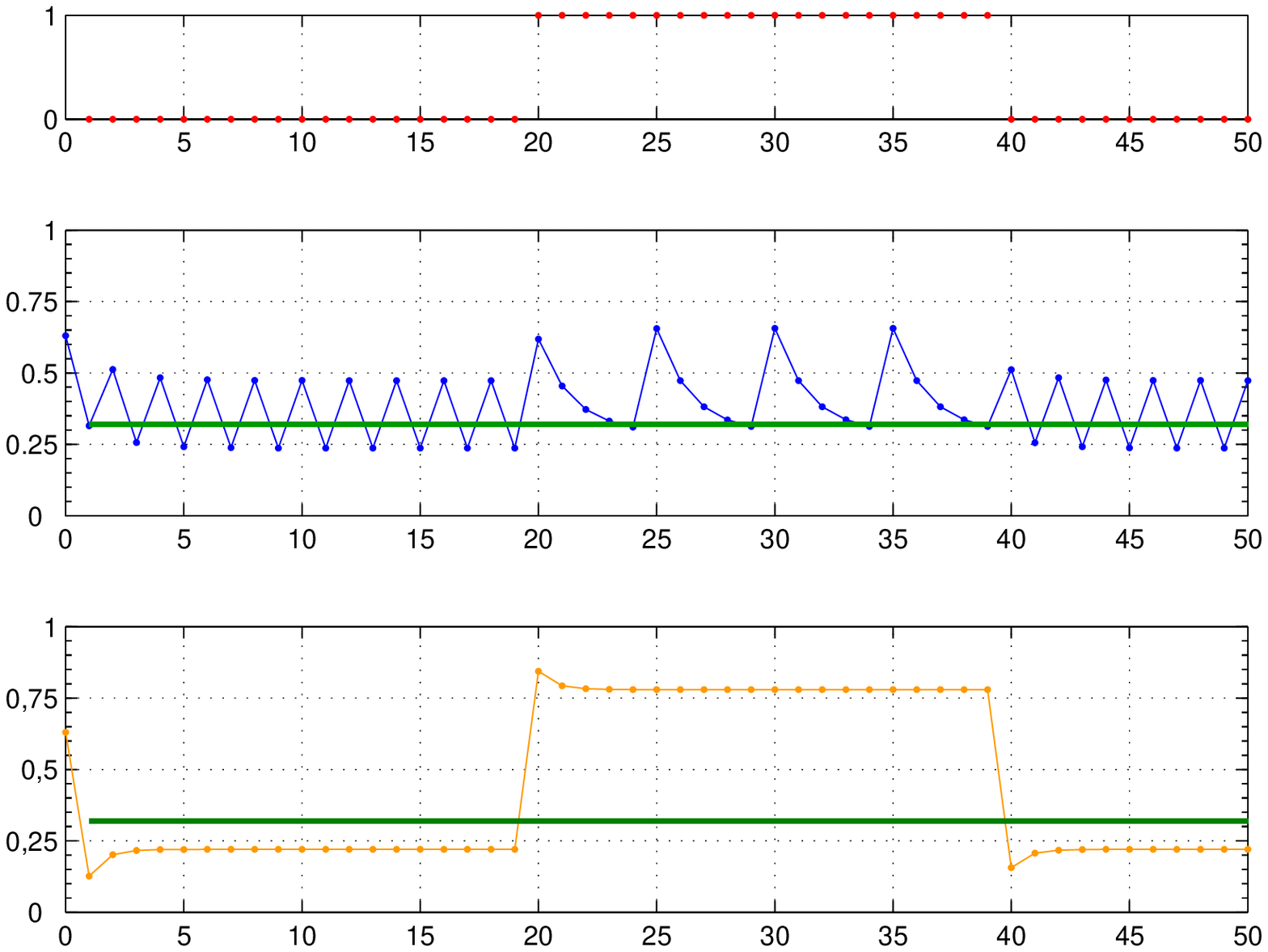}}
\put(180,194){\small {\bf (input)}} 
\put(60,180){{$\theta^t_u$}}
\put(320,165){{$t$}}

\put(135,154){\small {\bf (response case Figure 7 (A))}} 
\put(60,140){{$\x^t_v$}}
\put(320,90){{$t$}}

\put(135,80){\small {\bf (response case Figure 7 (C))}} 
\put(60,68){{$\x^t_v$}}
\put(320,15){{$t$}}
\end{picture}
\caption{\small {\bf The open self-inhibition}: The typical response $\x^t_v$ 
to the input $\x^t_u$: (A) in Region I,  small input intensity $\kappa_{uv}$, 
(C) in Region III,  high input intensity. The horizontal green line indicates 
the output threshold.}
\label{Figure_Self_inhibition_Region1_Region3}
\end{center}
\end{figure}

\ms Figure~\ref{Figure_Self_inhibition_Region1_Region3} displays the response of the 
self--inhibitor circuit to an input in Region I and Region III. The input $\theta^t_u$ 
is set to 0 for $0 \leq t<20$, then equal to 1 for $20 \leq t \leq 40$ and then again 
to 0 for $40<t<50$. Therefore, the dynamics is governed by the branches $f_{00}$ and 
$f_{10}$ during the first and the last period and by $f_{01}$ and $f_{11}$ during the 
intermediate time.
It is clear from Figure~\ref{Figure_Cuer_2} that the response is very different 
in case (A) of small input intensity and (C) of high input intensity.
In the first case there are no available fixed points and the system runs in pure 
oscillations and only the amplitude and the frequency distinguish the lower from the 
upper level of the input. In the last case the system contracts to the lower fixed 
point (see $f_{10}$) for lower level activation and to the upper fixed point 
(see $f_{01}$) for high level input. It is clear in this example that the input intensity 
may change the dynamical behavior of a module from oscillatory to bistable.

\ms 

\begin{figure}[h]
\setlength{\unitlength}{1pt}
\begin{center}
\begin{picture}(600,380)
\put(0,75){\includegraphics[scale=0.35]{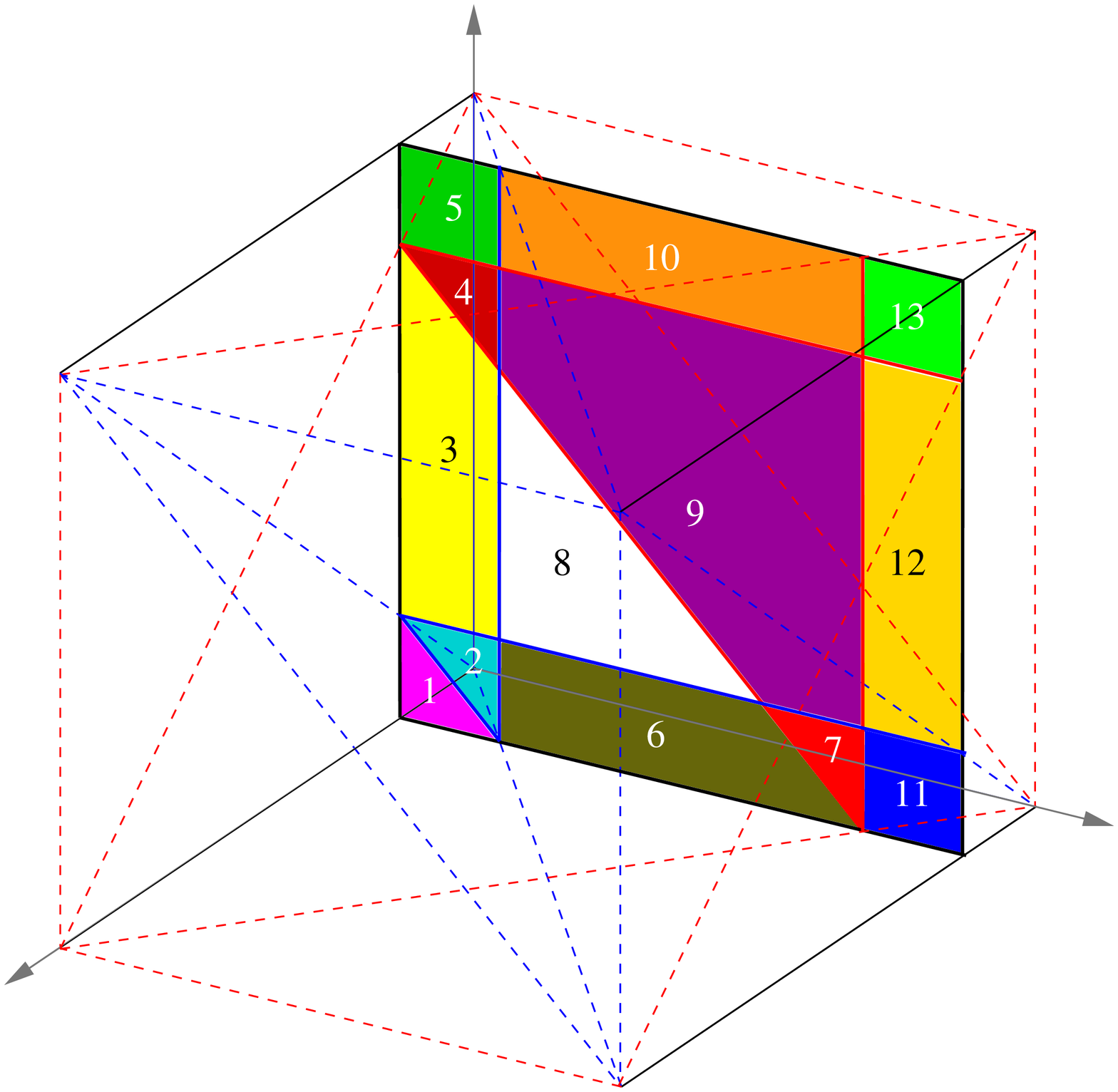}}
\put(215,118){{ $a$ }}
\put(-5,85){{ $T_{vv}$ }}
\put(70,290){{ $\kappa_{uv}$ }}
\put(130,0){{ (A) }}
\put(260,15){\includegraphics[scale=0.32]{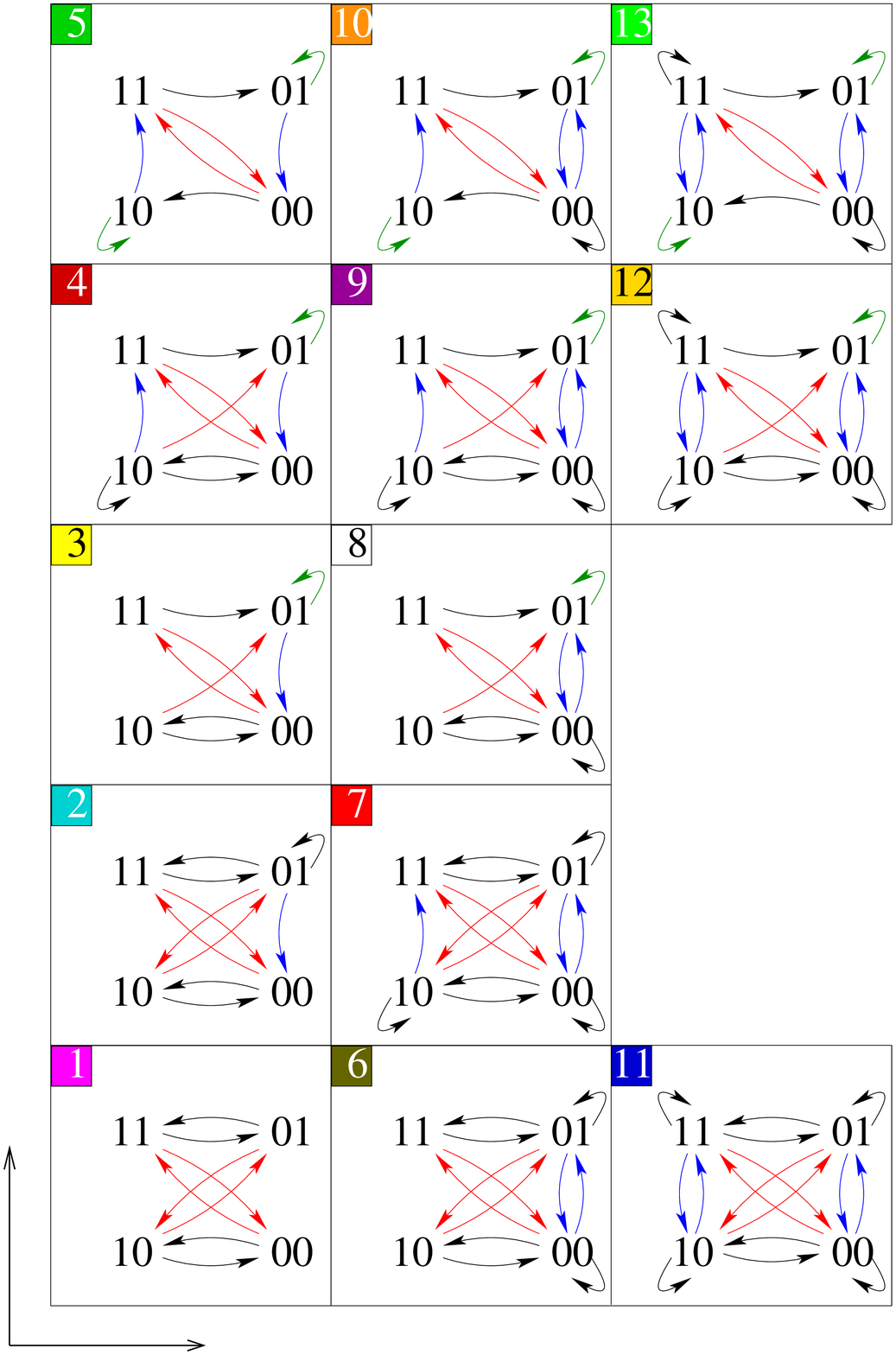}}
\put(295,6){{ $a$ }}
\put(240,50){{ $\kappa_{uv}$ }}
\put(350,0){{ (B) }}
\end{picture}
\end{center}
\caption{\small For the open self-inhibition, (A) Parameter space. It is made of 13 
sub--regions. Different colors, depicted for a given $T_{vv}$, show how the $\kappa_{uv}$ 
parameter can affect the dynamics. (B) Corresponding dynamical graphs of possible 
transitions. The code denotes $\theta_{vv} \theta_{uv}$. Colours are in correspondence 
with the plane in (A). {\bf \em Region I:} includes 1, 2, 6, 7, 11,  {\bf \em region II:}  
3, 4, 8, 9, 12 and {\bf \em region III:}  5, 10, 13.}~\label{OPEN_NEG_1D}
\end{figure}

\ms In Figure~\ref{OPEN_NEG_1D}, these three regions of parameters are subdivided 
in smaller subregions corresponding to different dynamical graphs and therefore to 
possibly different dynamical regimes that we describe in the following.

\ms 

\ms {\bf \em 1.- Region I.} For $\kappa_{uv} < T_{vv}$, corresponding to 
Figure~\ref{Figure_Cuer_2} (A), the dynamics is the same as for the isolated 
self--inhibitor, {\it i.~e.}, $\x_v$ oscillates whatever the input sequence 
can be, but with an amplitude and a frequency that depend on the input 
$(\theta_{uv}^t)_{t \in\mathbb{N}}$.
This is because in this case the corresponding maps have no fixed points.

\ms If $a$ is small enough (subregion 1), the system will change code at each 
time and the next code is completely defined by the external input at that time, 
Figure~\ref{OPEN_NEG_1D} (B).
In particular, for a constant input sequence the coding of the oscillating orbits 
$(\x_v^t)_{t \in \mathbb{N}}$ does not depend on the initial condition $\x_v^0$.

\ms For higher $a$ values (subregions 2, 6, 7 and 11), the input sequence 
$(\theta_{uv}^t)_{t \in \mathbb{N}}$ can non-trivially affect the dynamics of $x_v$.
This corresponds to the occurrence of the loops in the dynamical graph, allowing 
the orbit to lie inside a given atom of the symbolic partition more than one time 
step depending of the input sequence. The reason is clear:  as $a$ increases the 
image of the branches (see Figure~\ref{Figure_Cuer_2} (A)) may intersect the two 
sides of the discontinuity.

\ms 

\ms \begin{note}[The general case: local fixed points and absorbing intervals]
\begin{rm}
\ms One can consider the more general case of any number of inputs.
Again the existence of local fixed points for the IFS $F_{\theta_{uv}}$ depends on the 
internal and external parameter values.
Let $P$ be the number of local fixed points for $F_{\theta_{uv}}$.
Let $S_r(v)$ be any subset (possibly empty) of $I(v)$ such that 
$\sum_{u \in S_r(v)}  \kappa_{uv} > T_{vv}$ and let $R$ be the number of such subsets.
Let $S_l(v)$ be any subset (possibly empty) of $I(v) \cup \{v\}$ containing 
$\{v\}$ such that $\sum_{u \in S_l(v)}  \kappa_{uv} \leq T_{vv}$ and $L$ be the 
number of such subsets.
It is not difficult to see the relation: $P = L + R$.
Consequently $0 \leq P \leq 2^{\#I(v)}$. \\ \indent
Similarly, let $A$ be the number of local absorbing intervals $\subset (0,1)$ and 
let $S_c(v)$ be any subset (possibly empty) of $I(v)$ and 
$\bar{S}_c(v) = S_c(v) \cup \{v\}$ such that $\sum_{v \in S_c(v)} 
\kappa_{uv} \leq T_{vv} < \sum_{u \in \bar{S}_c(v)} \kappa_{uv}$, and $C$ be the 
number of such subsets.
Then $A = C$.
Consequently $0 \leq A \leq 2^{\#I(v)}$. \\ \indent
Finally one can check that $P + A = L+R+C = 2^{\#I(v)}$.
The two extrem cases are $P=0 \Leftrightarrow A=2^{\#I(v)}$ if and only if 
$1-\kappa_{vv} \leq T_{vv} < \kappa_{vv}$, and 
$P=2^{\#I(v)} \Leftrightarrow A=0$ if and only if 
$\kappa_{vv} \leq T_{vv} < \min_{u \in I(v)} \{\kappa_{uv}\}$ or 
$K_{vv} + \max_{u \in I(v)} \{\kappa_{uv}\}\leq T_{vv} < 1 - \kappa_{vv}$.

\ms Externally induced switches from (projected) fixed point converging regimes 
to (projected) periodic orbit converging regimes, and vice versa, give fairly 
simple dynamics.
It results in the concatenation of pieces of (transient) orbits from either regimes.
Switches between periodic orbit converging regimes may however be quite 
complicated as we see it next.
\end{rm}
\end{note}

\ms 

\ms {\bf \em 2.- Region III.} For the opposite case, i. e. for 
$\kappa_{uv} > 1-T_{vv}$ corresponding to Figure~\ref{Figure_Cuer_2} (C), 
we notice the occurrence of two fixed points.
The first, denoted $\x_{0}$, corresponds to the branch $f_{10}$ and therefore 
an orbit will come close to it by the repeated injection of the corresponding 
input $\theta_{uv} = 0$.
The second one, $\x_{1}$, corresponds to the branch $f_{01}$ and an orbit will 
came close to it by the repeated injection of the corresponding input 
$\theta_{uv} = 1$.
Therefore, in this case, the internal code $\theta_{vv}$ is a delayed slave of 
the input code $\theta_{uv}$ provided the forcing is permanent enough.

\ms In this sense the NAR behaves in this region of parameters as the RC described 
in the previous section.

\ms In particular, if $(\theta_{uv}^t)_{t \in \mathbb{N}}$ is constant 
($t$-independent), then there exists a $t^0 \geq 0$ such that 
$\theta_{vv}^t = 1 - \theta_{uv}^t \; \forall t \geq t^0$, so that the system 
can be easily driven in one of the two different states as if it was bistable.

\ms Notice that in the dynamical graphs of Figure~\ref{OPEN_NEG_1D} (B), for the 
corresponding subregions 5, 10 and 11, whatever the input sequence is, the loops 
that do not correspond to codes $10$ or $01$ cannot consecutively be taken an 
infinite number of times and the (finite) number of residence steps will depend 
on the input sequence after some delay.

\ms 

\ms If the input sequence $(\theta_{uv}^t)_{t \in \nn}$ is not constant, then the 
internal code $(\theta_{vv}^t)_{t\in \mathbb{N}}$ will depend on the initial 
condition $\x_v^0$.
In this case, depending of the parameters, the proper mathematical study of the 
internal code sequences that correspond to a given input sequence is still an open 
problem.

\ms The converse statement can however be formulated and is also of practical interest.
Namely, being given an observed internal sequence, what are the possible inputs 
and internal parameter values of the self-inhibition that realize that observed 
sequence?

\ms We now present an example showing that the use of {\it admissibility conditions} as 
in~\cite{CFML06},  implemented in a numerical algorithm, allows to produce such 
sequences for each particular occurrence.

\ms
\begin{example}\label{example3}
\begin{rm}
Let us illustrate the identification procedure in the case of a periodic orbit of an open 
self--inhibition with the observed internal sequence $(01001)^{\infty}$ of period 5.

\ms First, we indentify within Figure~\ref{OPEN_NEG_1D} (B) the candidate dynamical graphs 
to realize the observed sequence, and then the families of input sequence provided by 
those graphs.
In our example the candidate input sequences $(\theta^t_{uv})_{t \in \zz}$ are the 
families (with $\omega \in \{0,1\}$):
\begin{itemize}
 \item $(\omega \omega 1 \omega \omega)^\infty$ for $\fbox{2}$,
 \item $(0 \omega 1 0 \omega)^\infty$ for $\fbox{3}$ and $\fbox{4}$,
 \item $(0 1 1 0 1)^\infty$ for $\fbox{5}$,
 \item $(\omega \omega \omega \omega \omega)^\infty$ for $\fbox{6}$, $\fbox{7}$ and $\fbox{11}$,
 \item $(0 \omega \omega 0 \omega)^\infty$ for $\fbox{8}$, $\fbox{9}$ and $\fbox{12}$,
 \item $(0 1 \omega 0 1 )^\infty$ for $\fbox{10}$ and $\fbox{13}$.
\end{itemize}
Notice that no transitions in the graph $\fbox{1}$ can produce the observed internal code.

\ms Then we choose one input sequence among the more robusts.
Those are the one appearing in the greatest number of dynamical graphs.
This criterion makes more likely the set of parameter values that realize 
the observed internal code for that candidate input code to be broad in 
the parameter space.
This way we select the candidate sequence $(01101)^{\infty}$ common to all 
transition graphs except $\fbox{1}$, and for comparison we also consider 
$(00111)^{\infty}$ that is only possible for $\fbox{2}$, $\fbox{6}$, $\fbox{7}$ 
and $\fbox{11}$.

\ms Thirdly, we explicit the admissibility condition with the internal and 
the candidate input codes, and solve it.
The admissibility condition writes:
\begin{equation}\label{admissibility condition}
\sup_{t \in \zz : (\theta^t_{uu},\theta^t_{uv}) \in \{1\} \times \{0,1\} } \x^t
\leq T_{uu} \lesssim 
\inf_{t \in \zz : (\theta^t_{uu},\theta^t_{uv}) \in \{0\} \times \{0,1\} } \x^t,
\end{equation}
with for a fixed $(a,\kappa_{uv}) \in [0,1) \times [0,1]$,
\begin{equation}\label{orbit}
\left\{ \x^t \;:\; t \in \zz \right\} = 
\left\{ 1-\frac{1-a}{1-a^5}\left[ (1-\kappa_{uv}) 
\displaystyle{\sum_{k \in I_s}} a^{(n+k) \text{mod} 5} + 
\kappa_{uv} \displaystyle{\sum_{k \in E}} a^{(p+k) \text{mod} 5} \right] 
\;:\; (n,p) \in \llbracket0,4\rrbracket^2  \right\}.
\end{equation}
In the last expression $I_s = \{ k \in \llbracket0,4\rrbracket
 \;|\; \sigma^{-1}(\theta_{uu})=0 \text{ and } \sigma^{-k-1}(\theta_{uu})=0 
 \text{ if } s=0 \text{ and } \sigma^{-k-1}(\theta_{uu})=1 \text{ if } s=1 \}$, and
$E = \{ k \in \llbracket0,4\rrbracket \;|\; \sigma^{-1}(\theta_{uv})=0 
\text{ and } \sigma^{-k-1}(\theta_{uv})=0 \}$, $\sigma$ being the left 
shift map in $\{0,1\}^{\zz}$.
We have explicitely in our case $I_0=\{1,2,4\}$, $I_1=\{0,3\}$, and $E=\{2,4\}$.

\ms The results are shown Figure~\ref{IDENTIFICATION} with the plot of all the 
$(a,T_{vv})$ values admissibles for a fixed value of $\kappa_{uv}$.
As expected from the analysis of the transition graphs, the domain for 
$(00111)^{\infty}$ appears smaller than the domain for $(01101)^{\infty}$ 
when varying the $\kappa_{uv}$ parameter.

\ms 
\begin{figure}[h]
\setlength{\unitlength}{1pt}
\begin{center}
\begin{picture}(550,200)
\put(15,0){\includegraphics[scale=0.25]{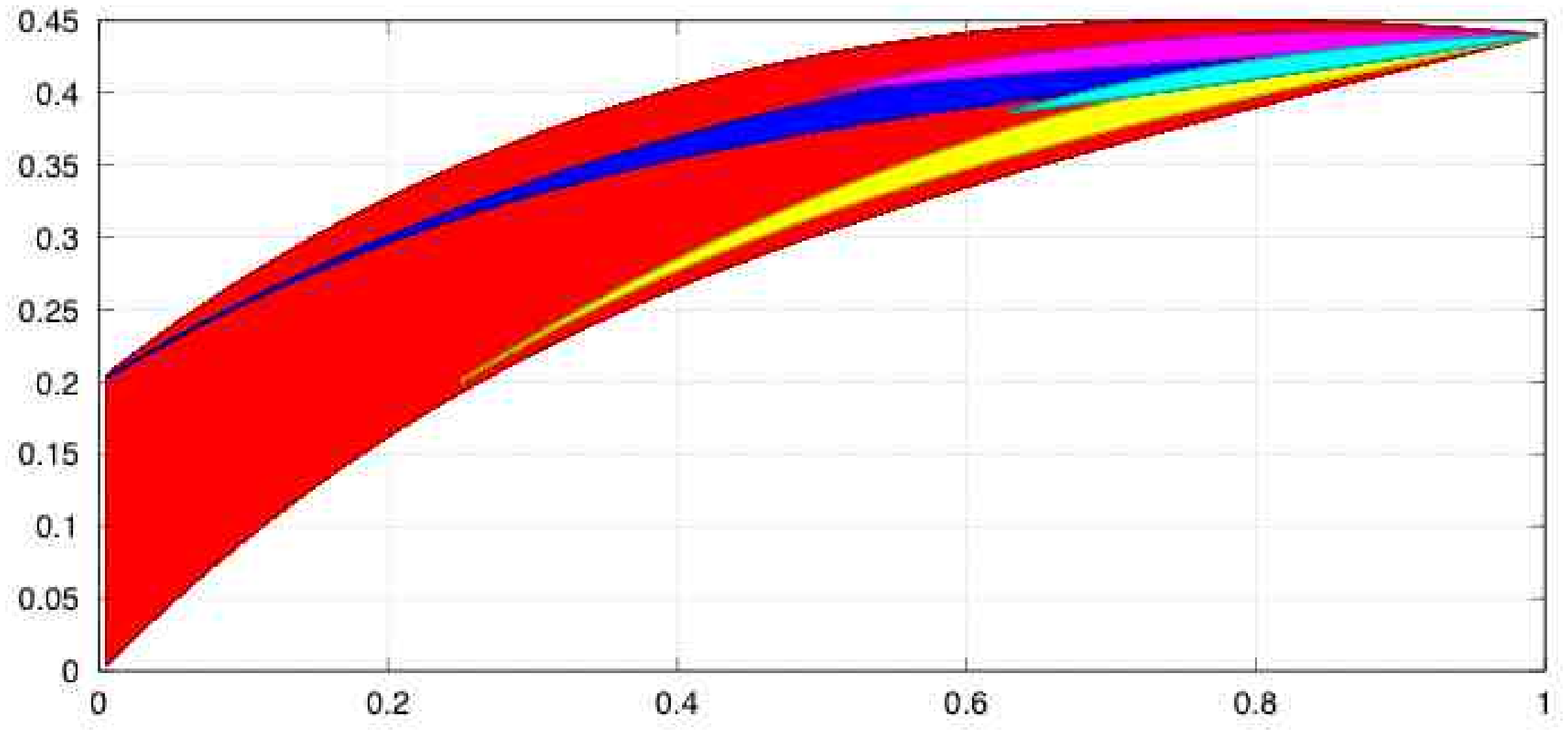}}
\put(217,17){{ $a$ }}
\put(0,123){{ $T_{vv}$ }}
\put(110,0){{ (A) }}
\put(15,90){\includegraphics[scale=0.25]{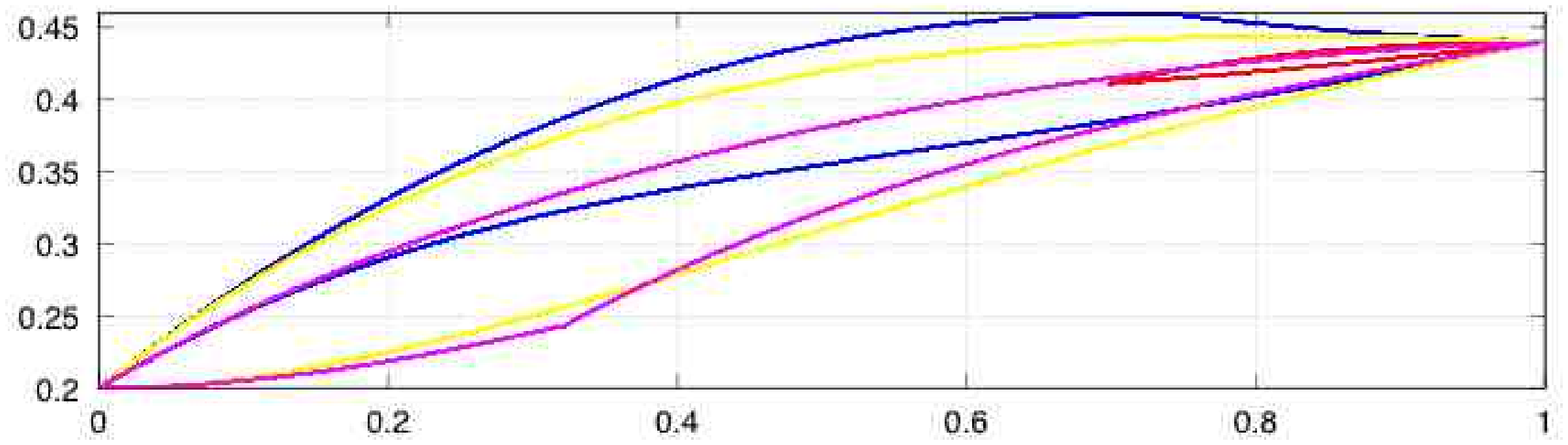}}
\put(110,195){{ (B) }}
\put(230,7.5){\includegraphics[scale=0.25]{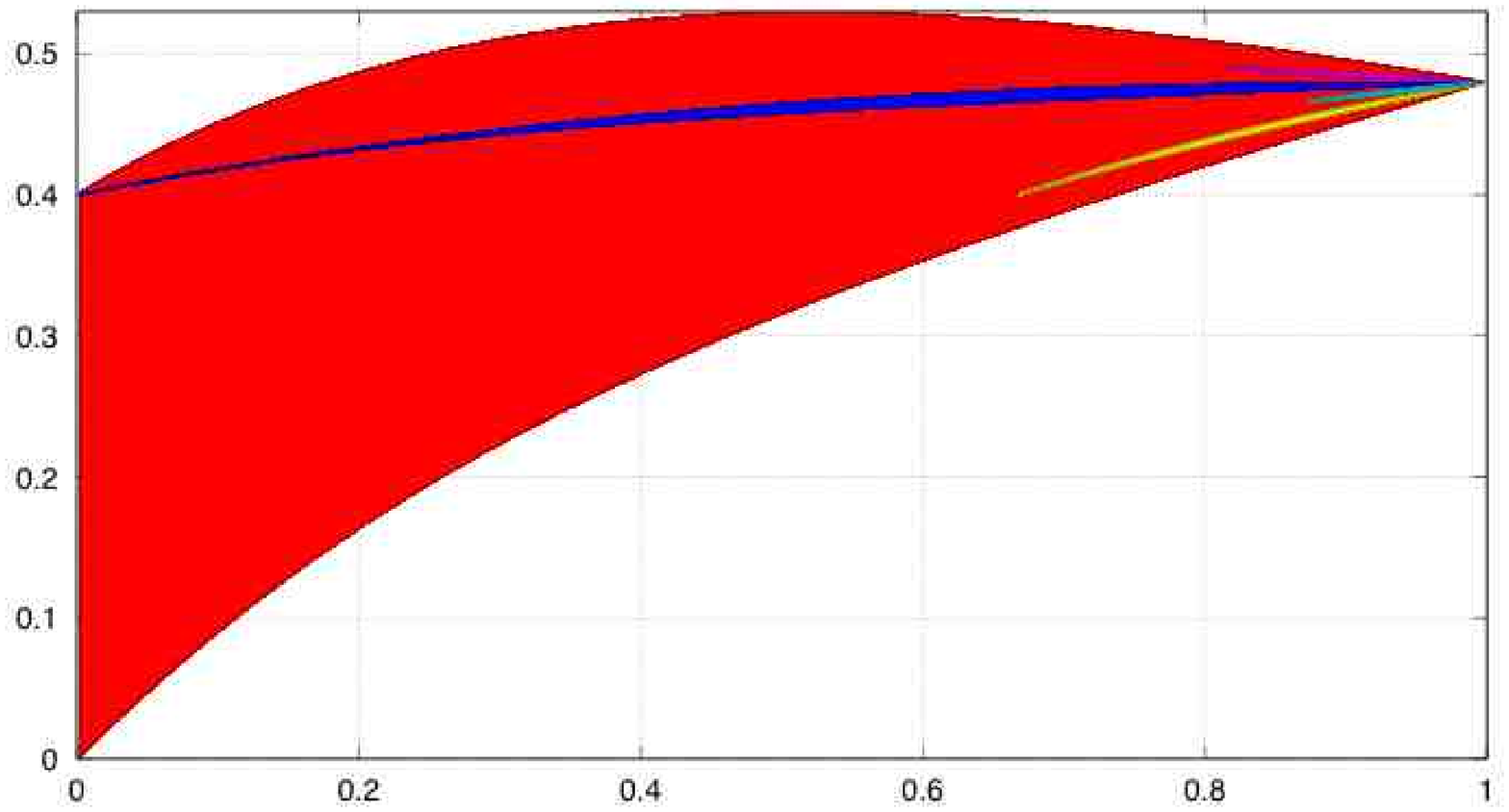}}
\put(325,0){{ (C) }}
\put(228,101){\includegraphics[scale=0.25]{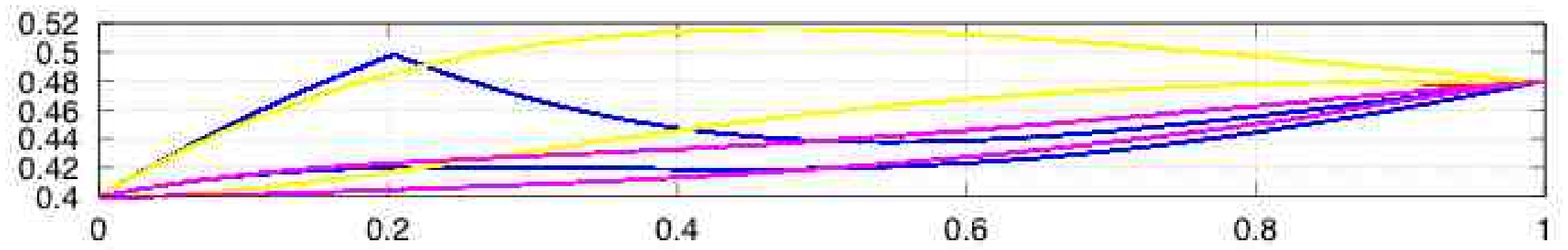}}
\put(325,195){{ (D) }}
\end{picture}
\end{center}
\caption{\small $(a,T)$ subdomains of admissibility for the periodic input 
$(01101)^{\infty}$ with $\kappa_{uv}=0.2$ (A) and $\kappa_{uv}=0.4$ (B), 
and for the periodic input $(00111)^{\infty}$ with $\kappa_{uv}=0.2$ (C) 
and $\kappa_{uv}=0.4$ (D). For (A) and (C) the admissible values are within 
the colored areas while for (B) and (D) the admissible values are in 
between the colored boundaries. The area and boundary colors are in 
correspondance.}~\label{IDENTIFICATION}
\end{figure}

\ms One can see the overlap of the domains for a given value of $K_{uv}$ 
(Compare (A) and (B) for instance), meaning that the same observed internal 
sequence can be observed for the same parameter values for different 
input sequences. It illustrates the robustness to that parameter.

\ms Also, depending on the input sequence there may have a dependence to the 
initial condition. This is illustrated on Figure~\ref{IDENTIFICATION} where each 
color corresponds to an attracting periodic orbit with its bassin of attraction.
Furthermore, although all the attracting periodic orbits may have the same period, 
they differ in particular by their amplitude.
Another way to present it is that in pratice different realizations may 
look similar (frequency) but differ (in the state at a given time, in amplitude) 
depending on the initial condition.

\ms Finally, notice the qualitative relation between the domains of 
Figure~\ref{IDENTIFICATION} and those of Figure~\ref{OPEN_NEG_1D} (A).
The sequence $(01001)^{\infty}$ is indeed for the input $(01101)^{\infty}$ out of 
the tetrahedron $\fbox{1}$ as indicated by the derivative of the uppen 
bound of the red domain a $a=0$ Figure~\ref{IDENTIFICATION} (A).
Also The sequence $(01001)^{\infty}$ is indeed for the input 
$(00111)^{\infty}$ out of the tetrahedron $\fbox{8}$.
\end{rm}
\end{example}

\ms 
\begin{note}
\begin{rm}
\ms System identification is closely related to the control issue, as if 
we can find out an input and parameter values that realize a given behavior 
of the module, then that input, with the appropriate parameter values, can 
also be used to control the module to a desired behavior. Furthermore the 
robustness to small aditional perturbations has been illustrated on 
Figure~\ref{IDENTIFICATION}.
\end{rm}
\end{note}

\ms {\bf \em 3.- Region II.} We first treat the case $1- \kappa_{uv} < \kappa_{uv}$.

\ms If $T_{vv} \leq \kappa_{uv} < 1-T_{vv}$, corresponding to 
Figure~\ref{Figure_Cuer_2} (B), the dynamics shares the characteristics of both previous systems.
This is because in this parameter region only the branch $f_{01}$ (matching the 
input  $\theta_{uv} = 1$) induces a 
fixed point, while on the branch $f_{00}$ the system will oscillate in any case.
That is to say: if $\theta_{uv}^t = 1\ \forall t\in \nn$, then any orbit converges 
asymptotically to $\x_{1}$, if $\theta_{uv}^t =0\ \forall t\in\nn$, then the dynamics 
is as for the pure self--inhibition, up to a change of variable that depends on 
$\kappa_{uv}$ (the same as for region I).

\ms The region II contains three subregions denoted 3, 4, 8, 9 and 12 in in 
Figure~\ref{OPEN_NEG_1D}. Again, depending on the subregion, some transitions, in 
presence of a suitable input, are possible or not.
As in the previous case, and for the same reason, only the loop corresponding to $01$ in 
the dynamical graph may be taken an infinite number of consecutive time steps.
Again, the (finite) number of residence steps in the remainder codes will depend on the 
input sequence after some delay.

\ms In case that  $ \kappa_{uv} < 1- \kappa_{uv}$, due to the symmetry 
$\kappa_{uv}\longleftrightarrow 1-\kappa_{uv}$ we shall have in Region II a steady low 
level state $\x_{0} < T_{vv}$ corresponding to the input $\theta_{uv}^t = 0$ and 
high level oscillation corresponding to the input  $\theta_{uv}^t = 1$.

\ms Finally, we see that depending on the parameter values of the circuit and on the 
different inputs the circuit may either oscillations, exhibit a low level or a high level 
steady state, or even be bistable.

\ms This analysis shows that the range of external signals can have a significant 
influence on the expression dynamics of a self--inhibited gene. Depending on the input 
signal and/or the self--regulation parameters, very different dynamical regimes can 
exist and satisfy different functional 'demands' of the whole regulatory network 
through the subsequent interactions the gene is involved in.

\ms \subsubsection{\bf The open self--activation}\

\ms We now proceed with the positive self-regulation (PAR).
An isolated PAR is a bistable system apart from the (non functional) case where 
$T_{vv} \notin [0,1] $ \cite{CFML06}.

\begin{figure}
\setlength{\unitlength}{1pt}
\begin{center}
\includegraphics[scale=0.3]{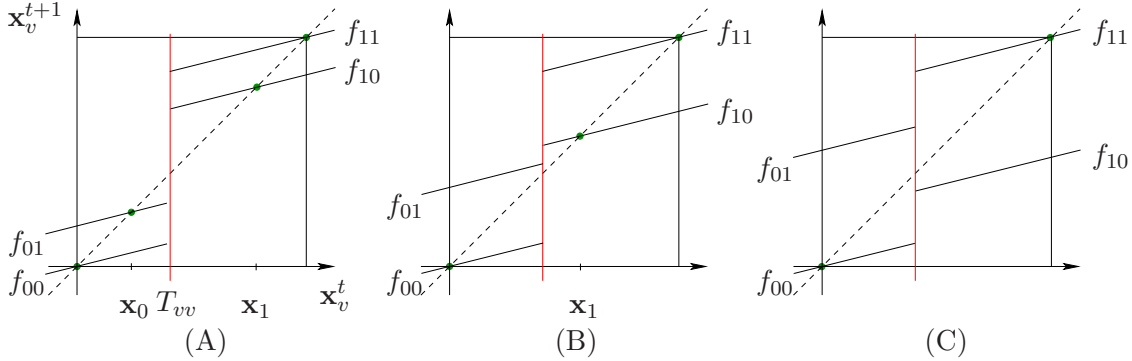}
\begin{picture}(400,0)
\put(60,-5){\makebox(10,0){(A)}}
\put(33,8){\makebox(10,0){$\x_{0}$}}
\put(49,10){\makebox(10,0){$T_{vv}$}}
\put(79,8){\makebox(10,0){$\x_{1}$}}
\put(108,10){{$\x^{t}_{v}$}}
\put(-10,115){{$\x^{t+1}_{v}$}}
\put(-10,30){{$f_{01}$}}
\put(-10,15){{$f_{00}$}}
\put(117,95){{$f_{10}$}}
\put(117,110){{$f_{11}$}}

\put(200,-5){\makebox(10,0){(B)}}
\put(203,8){\makebox(10,0){$\x_{1}$}}
\put(133,44){{$f_{01}$}}
\put(133,15){{$f_{00}$}}
\put(258,80){{$f_{10}$}}
\put(258,110){{$f_{11}$}}

\put(340,-5){\makebox(10,0){(C)}}
\put(273,58){{$f_{01}$}}
\put(273,15){{$f_{00}$}}
\put(400,63){{$f_{10}$}}
\put(400,110){{$f_{11}$}}
\end{picture}
\caption{\small {\bf The open self--activation.} Graphs of the IFS
$F_{\theta_{uv}}(\x_{v}) = a \x_{v} +(1-a) 
 \left[ H \left (\x_v-T_{vv} \right) + \kappa_{uv} 
  \left( \theta_{uv} - H \left (\x_v-T_{vv} \right) \right) \right].
$
For (A) Region I, small input intensity, (B) Region II, intermediate input intensity and (C) Region 
III, high input intensity. Possible fixed points are given by the intersection of the graph with 
the diagonal. The notation $f_{ij}, i,j=0,1$ stands for the branch of $F_{\theta_{uv}}$ when 
$H \left(\x_{v}-T_{vv}\right)=i$ and $\theta_{uv}=j$.}
\label{supplementary_Cuer_6}
\end{center}
\end{figure}

\ms Figure~\ref{supplementary_Cuer_6} illustrates the three possible cases of the 
dynamics for the corresponding open system, shown Figure~\ref{supplementary_Cuer_1}, 
when varying the input intensity. Accordingly, as in the case of the self-inhibition, 
it follows that the parameter subspace 
$\{ (T_{vv}, a, \kappa_{uv}) \;:\; T_{vv} \in (0,1), a \in [0,1) \; \text{and} \; 
\kappa_{uv} \in [0,1] \}$
can be divided into three input intensity regions corresponding to different dynamical 
characteristics. The corresponding parameter domains are as for the NAR: 
{\bf \em region I:} if $\kappa_{uv} < T_{vv}$, {\bf \em region II:} if 
$T_{vv}<\kappa_{uv} < 1-T_{vv}$ and {\bf \em region III:} if $\kappa_{uv} > T_{vv}$.

\ms Depending of the input intensity, some or all of the possible fixed points are present:
$0$ and $\x_{0}$ corresponding to the input $\theta_{uv} = 0$  ($0$ is present for all values of 
the input intensity) and  $(\x_{1}$ and $1$ corresponding the input, $\theta_{uv} = 1$ 
(see Figure~\ref{supplementary_Cuer_6}). Note that $1$ is present for all values of the 
input intensity.

\begin{note} 
\begin{rm} We will consider only the cases were $T_{vv} < 1-T_{vv}$ for the same reason than for 
the open self--inhibition.
\end{rm}
\end{note}

\ms 


\ms
{\bf \em 1.- Region I.} For $\kappa_{uv} < T_{vv}$, corresponding to 
Figure~\ref{supplementary_Cuer_6} (A), there exists two absorbing regions, 
$(0,\x_{1})$ and $(\x_{0}, 1)$, for 
the orbits $(\x_v^t)_{t \in \mathbb{N}}$ on each side of the threshold $T_{vv}$ 
whose size is fixed by the input intensity $\kappa_{uv}$.
(Notice that if $\kappa_{uv} = 0$ then the dynamics is that of the isolated self-activation.)

\ms Therefore, for a constant input sequence the dynamics is bistable and an orbit 
$\x_v^t$ converges to either $0$ or $\x_{0}$, depending on the initial condition, if 
$\theta_{uv}^t = 0 \; \forall t$, or to either $\x_{1}$ or $1$ if 
$\theta_{uv}^t = 1 \; \forall t$.

\ms If $\theta_{uv}^t $ is not constant, then the dynamics of $\x^t_v$ is driven by the 
input sequence, with possible oscillations within one of the two absorbing intervals 
$(0,\x_{1})$ and $(\x_{0}, 1)$.

\ms {\bf \em 2.- Region III.} For $\kappa_{uv} \geq 1-T_{vv}$ , corresponding to 
Figure~\ref{supplementary_Cuer_6} (C), whatever the value of $a$, if the input 
sequence is constant then the dynamics has one fixed point, at $0$ if $\theta_{uv} = 0$ 
and at $1$ if $\theta_{uv}= 1$.

\ms In this input intensity region the open self-inhibition is a slave to the input 
sequence. It exhibits, however, a faster convergence to either fixed point if the input 
intensity $\kappa_{uv}$ is close to $T_{vv}$, ~\cite{AlonNature2007}.

\ms As for the self--inhibition in the same range of parameters, if the input sequence 
$(\theta_{uv}^t)_{t \in \mathbb{N}}$ is not constant, then the internal code 
$(\theta_{vv}^t)_{t\in \mathbb{N}}$ will depend on the initial condition 
$\x_v^0$ and again, using the {\it admissibility conditions} as in~\cite{CFML06}, it is 
possible to produce such sequences in each particular (see Example~\ref{example3}).

\ms {\bf \em 3.- Region II.}
We start with the case $1 - \kappa_{uv} <  \kappa_{uv}$.

\ms If $T_{vv} \leq \kappa_{uv} < 1-T_{vv}$ (see Figure~\ref{supplementary_Cuer_6} (B)), 
depending on the input sequence $(\theta_{uv}^t)_{t \in \mathbb{N}}$, the dynamics can be 
either monostable,  with attracting fixed point $1$ if $\theta_{uv}^t = 1 \; \forall t$, 
bistable with two possible attracting fixed points $0$ and $\x_0$ depending on initial 
conditions if $\theta_{uv}^t = 0 \; \forall t$, or showing more complicated (oscillating) 
dynamics if $\theta_{vv}^t$ is not constant.

\ms As for the open self-inhibition circuit if $ \kappa_{uv} < 1- \kappa_{uv}$, due to 
the symmetry $\kappa_{uv}$ $\longleftrightarrow$ $1- \kappa_{uv}$  the dynamics in 
Region II can be either monostable, with attracting fixed point $0$ if 
$\theta_{uv}^t = 0 \; \forall t$, or bistable with two possible attracting fixed points 
$1$ and $\x_1$ depending on initial conditions if $\theta_{uv}^t = 1 \; \forall t$.

\ms Finally we see that, depending on the parameter values of the circuit and on the different 
inputs, the circuit may either oscillate, or be bistable for both inputs, or to show a low 
level or a high level steady state and bistability.

\ms This classification of the different dynamical regimes may be refined exactly in the 
same way as for the open self-inhibitor. It is clear from the analysis made for the NAR 
that crossing in the parameter space the same limits as for the self-inhibitor will cause 
robust bifurcations, leading to changes in the corresponding dynamical graph.

\ms Finally, as for the self-inhibition, this analysis shows that the range of external 
signals can have a significant influence on the expression dynamics of a self-activated gene.
Depending on the input signal and/or the self-regulation parameters very different dynamical 
regimes can exist.

\ms \subsection{The open negative  2--circuit}\

\ms
The dynamics of the negative 2--circuit is studied in detail in \cite{CFML06}.
It has a very reach dynamics including periodic and quasi-periodic attractors that may 
coexist for certain parameter values and there are no fixed points.
These attractors organize the dynamics in the phase space. We shall describe the main 
features of the dynamics of the 
open negative 2--circuit shown in Figure~\ref{P53-MDM2}, as compared with the autonomous system.

\ms 
\begin{center}
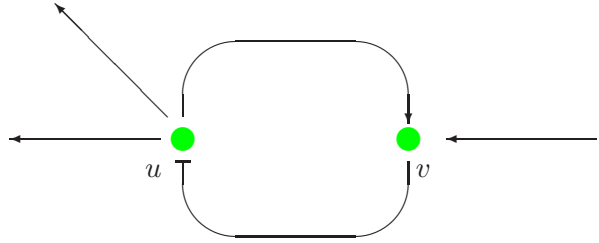
\begin{figure}[h]
\setlength{\unitlength}{1truecm}

\begin{picture}(8,4)(0,0)
\put(2,2){\color[rgb]{0,1,0}\circle*{0.3}}
\put(5,2){\color[rgb]{0,1,0}\circle*{0.3}}
\put(7.5,2){\vector(-1,0){2}}
\put(1.8,2.3){\vector(-1,1){1.5}}
\put(1.7,2){\vector(-1,0){2}}
\put(3.5,2.3){\oval(3,2)[t]}
\put(3.5,1.7){\oval(3,2)[b]}
\put(5,2.4){\vector(0,-1){0.2}}
\put(1.9,1.7){\line(1,0){0.2}}
\put(1.5,1,5){$u$}
\put(5.1,1.5){$v$}

\end{picture}
\ms
\caption{\small The open negative 2--circuit.}\label{P53-MDM2}
\end{figure}
\end{center}


\ms
According to the last comment in section~\ref{Open_networks}, it is important to 
understand first the cases where the forcing sequences are constant, 
$\theta_{\rm in}^t = 0$ or $\theta_{\rm in}^t = 1$ for all $t$.
For $\theta_{\rm in}^t = 0$ and $1$, consider the two maps $F_{\theta}$ as 
in~\eqref{Fteta} and 
denote $\x_{\theta}$ their fixed points different from $0$ and $1$.

\ms There are two completely different cases for that system.
 
\ms First, {\it the oscillatory induced regime} that works as for the autonomous 
negative circuit: if  $T_{vv} < \x_0$ and for $0$ as input, any trajectory of the 
$\x_v$ component of $F_0$ will end up after a finite number of time steps inside the 
invariant interval $[0, \x_0)$ and from there, up to an affine change of variables, the 
dynamics of $F_0$ behaves as the negative autonomous 2--circuit studied in \cite{CFML06}: 
i. e. it oscillates.
The symmetric situation occurs for $\x_1 < T_{vv} $ and $1$ as input.
After a finite number of steps the dynamics of the $\x_v$ component of $F_1$ oscillates 
inside the interval $(\x_1, 1]$. In short, the two variables $\x_u$ and $\x_v$ oscillate 
out of phase (by $\pi /2$).
In case of $0$ forcing they oscillate at a low level and in case of $1$ as input at a high level.
This situation occurs after some transient time if the initial condition of $\x_v$ happen to 
be outside the corresponding invariant interval.

\ms Second, {\it the fixed induced regime} which is different from the autonous case: 
if $T_{vv} > \x_0$ and for $0$ as input, any trajectory will be attracted by the unique 
fixed point of $F_0$. This fixed point is $(1, \x_0)$.
The reason is that, after a finite time we get  $\x_v^t < T_{vv}$ and from then the signal 
$\theta_{vu}^t$ sent to node $u$ is always $1$ (since this interaction is an inhibition).
As consequences, $\x_u \rightarrow 1$ and the signal $\theta_{uv}^t$ sent to node $v$ is 
then always $1$ (since now this interaction is an activation).
Therefore, also $\x_v \rightarrow \x_0$.
Finally, in this case, for a $0$ as input, the circuit ends up in $\x_u= 1$, $\x_v=\x_0$.
The case $\x_1>T_{vv}$ and $1$ as input is solved in the same way, and the circuit ends up 
in $\x_u= 0$, $\x_v=\x_1$.

\ms Notice that in the fixed induced case, contrarily to the autonomous circuit, it is not the initial conditions that determines the final destination of the system but the external input.

\ms Now, by a convenient choice of the external versus internal intensity it is possible to set 
$\x_0 < \x_1$ or $\x_0 > \x_1$.
Therefore there are four possible open negative 2--circuits: (1) the bi--oscillating circuit 
that oscillates at small amplitude for a low level input and at a high amplitude for a high 
level input; (2) the oscillating--fixed circuit that oscillates at low amplitude for low 
level input and converges to $\x_u= 0$, $\x_v=\x_1$ for high level input; (3) the 
fixed--oscillating circuit converging to $\x_u= 1$, $\x_v=\x_0$ for low level input and 
oscillating for high level input and (4) the bistable circuit that converges to $\x_u= 1$, 
$\x_v=\x_0$ or to $\x_u= 0$, $\x_v=\x_1$ according to a low or high input level.

\ms Finally, for a more general (time variable) input this open circuit may work in different 
manners and, by appropriately tuning the forcing it is possible to switch the circuit from 
one regime to another.

\ms By the same arguments, the open negative 2--circuit with reversed sign of interactions works in the same manner.

\ms \subsection{The open positive 2--circuit}\ 

\ms The same type of argument shows that for a positive 2--circuit and a constant external 
forcing, in one case it is bistable as is the autonomous system \cite{CFML06}, but becomes 
monostable in the opposite case.
Therefore, it is also possible to build bistable--bistable, bistable--monostable, 
monostable--bistable and monostable--monostable circuits working in the pointed regimes for 
low--high input levels.

\ms Again it is clear in this case that an open positive circuit may operate in a regime 
different from the autonomous counterpart.
In particular, an oscillatory input may drive the positive circuit in oscillations.

\ms Notice that in~\cite{AMVM06} the same discrete--time piecewise--affine model have 
been numerically studied to account for the dynamics of a p53--Mdm2 genetic regulatory 
network made of an open positive and an open negative 2--circuit in interaction, with one 
inward regulation acting on each of the two circuits. In our notation, node $u$ stands 
for p53 and node $v$ for Mdm2.

\ms The analysis above extend to the case of open n-circuits, in the same way as for the
corresponding autonomous circuits \cite{CFML06}.

\bs\section{Example of mixed type}\

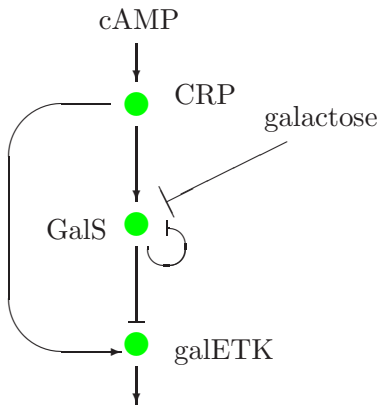
\begin{figure}[h]
\setlength{\unitlength}{1truecm}
\begin{center}
\begin{picture}(4,5)(0,0)

\put(2.5,5){cAMP}
\put(3,4.8){\vector(0,-1){0.5}}

\put(3,4){\color[rgb]{0,1,0}\circle*{0.3}} 
\put(3.5,4){CRP}
\put(3,3.7){\vector(0,-1){1}}

\put(3,2.4){\color[rgb]{0,1,0}\circle*{0.3}}
\put(1.8,2.2){GalS}
\put(3,2.1){\line(0,-1){1}}
\put(2.9,1.1){\line(1,0){0.2}}

\put(5,3.5){\line(-2,-1){1.6}}
\put(3.3,2.9){\line(1,-2){0.2}}
\put(4.7,3.7){galactose}

\put(3.4,2.1){\oval(0.5,0.5)[b]}
\put(3.4,2.1){\oval(0.5,0.5)[r]}
\put(3.4,2.25){\line(0,1){0.2}}

\put(3,0.8){\color[rgb]{0,1,0}\circle*{0.3}}
\put(3.5,0.6){galETK}
\put(3,0.5){\vector(0,-1){0.5}}

\put(2.65,2.35){\oval(2.7,3.3)[l]}
\put(2.6,0.7){\vector(1,0){0.2}}

\end{picture}
\ms
\caption{\small Incoherent feedforward loop of the type 1 
from~\cite[pag.~454]{AlonNature2007}.}~\label{figAlon}
\end{center}
\end{figure}

\ms In the more general case when a module owns one or more circuits inside a regulatory 
cascade it is in principle possible to combine our results and describe all its possible 
dynamical behaviors. For large networks this method will soon end up in the study of the 
network as a all and therefore it is of any help. However for simple networks the splitting 
of the all network in elementary RC and FC modules may help the study of their dynamics. 
As an example let us consider the incoherent type 1 feeforward loop 
(I1--FFL)~\cite{AlonNature2007}, depicted in Figure \ref{figAlon}.
We first determine the possible dynamical regimes for the open self--inhibitor present 
at the level of GalS.
Since there are two external inputs for this circuit (CRP and galactose) its dynamics 
displays four branches, corresponding to the possible combinations of the two inputs.

\ms

\ms 

\ms Therefore we can organize this information in a $2\times 2$ table where say, arrows 
correspond to $\theta_{CRP,GalS} = 0$ or $1$ and columns  to $\theta_{galactose} = 0$ or $1$ 
(notice that, because galactose repress GalS, according to our notation, 
$\theta_{galactose} = 0$ stands for high galactose level and $\theta_{galactose} = 1$ for 
low level). Now by looking for intensities of the interactions and thresholds, each entry 
of this table can be filled on the basis of the 3 possibilities described 
in~\ref{The open self-inhibition}: steady low level, steady high level or oscillation.
In each case this information immediately fixes $\theta_{GalS, galETK} ^ {t + \tau_2}$  
where $ \tau_2$ is the delay due to the interaction ${GalS, galETK}$ as described 
in~\ref{The general regulatory cascade}.
Notice also that $\theta_{CRP,GalS}^t$ is just equal to the input signal 
$\theta_{cAMP}^{t-\tau_1}$ after the delay  $\tau_1$ due to the node $CRP$.

\ms Of course the delays $\tau_1$ and $\tau_2$ are only important during a transient time 
after one of the external signals eventually changed (as it is the case for a pulse input).
If not, time translation invariance of the inputs simplify the analysis.

\ms It is then sufficient to incorporate this information in the description of the dynamics 
of the RC made in~\ref{The general regulatory cascade}. As a consequence this module can 
end up with, either an oscillatory regime for the final gene operon $galETK$ corresponding 
to an input corresponding to an ``oscillation'' entry of the table, or in a ``steady state'' 
(low or high level) for an input with this entry.

\ms
We emphasize that this module may operate in very different regimes either by a change of 
parameters (due for instance to a mutation) or by some modulation of the input signals, or both.

\bs
\section{Final comments}~\label{section-final} \

\ms
In this work we made an attempt to understand how a small regulatory network is operating 
under external stimulus. This stimulus can be, either the output of a larger network where 
the module is inserted or simply an external signal acting as a trigger mechanism for the 
action of the module.

\ms For Regulatory Cascades (RC) we have shown that there are two (non exclusive) conditions 
under which the RC acts as a finite time delayed transducer (or translator) exactly as a 
cellular automata. It takes a finite input ``message'' in another finite output ``message''.
This two conditions are, either an internal property of the module (readable in its 
parameter values) or an external characteristic of the input signal.
In a sense we can say that under one (or both) of this conditions the module works in a 
safety operating regime since it can transmit an unambiguous command with only some delay.

\ms In the opposite case, the situation is different since then, in principle, the module 
needs a variable, eventually infinite, time to ``understand'' the input message and then 
to be able to produce an unambiguous output. Moreover this time delay may depend of the 
particular input signal (see final part of Remark~\ref{long-remark}).
No doubt that in this case the module is not of a great help for the system, even if 
this behavior is fascinating from the point of view of nonlinear dynamics.
But maybe Nature has found already somewhere a situation where this fuzzy--like operating 
regime has some selective advantage!

\ms From a mathematical point of view our strategy was first to prove these properties on 
the simplest possible module (the elementary transducer ET) and then extend them step by 
step up to the general case (RC).

\ms
For the Forced Circuits (FC) we use a slightly different strategy.
In each case we work out the subsets of the phase space involved in the dynamics under each 
external input.
Here the case of stationary inputs enable a simple classification of the different 
``extreme'' dynamical responses of the module.
Then thanks to the contracting (or diffusive) properties of the dynamics (a consequence 
of degradation) we may predict the dynamics under more general inputs.
This also opens the possibility to build small modules with different desired functionalities.
From a mathematical point of view our strategy was simply to restrict the dynamics in each 
FC to the part of the phase space that became invariant and attractive under each specific input.

\ms 
These small open circuits show how their functioning may depend on the parameters as well 
as on the different type of inputs.
In this context, the notion of modularity for the accomplishment of a function cannot 
be reduced to a simple decomposition of the network into subsets of nodes and interactions.

\ms 
An example where such ambiguity may be interesting for a biological function is the 
case where for a while a cell population needs two types of differentiated cells 
(corresponding to low and medium expression level of genes for instance, see 
Figure~\ref{supplementary_Cuer_6} (B), up to a moment when, under an appropriate signal 
(generated by a stress for instance), the two states switch to the same effective 
differentiated state (high expression level of
genes for instance), see Figure~\ref{supplementary_Cuer_6} (C).

\ms
Now we may wonder why we need one point of view (transducer) when dealing with RC and 
another (constrained phase space) to study FC. The two points of view are in fact equivalent.
The bridge between them is symbolic dynamics. The starting point of symbolic dynamics is 
to encode the atoms of a suitable partition of the phase space by a set of 
symbols and to transfer the description of the dynamics in the phase space in 
terms of rules on the corresponding symbolic representation.
In our case the partition of phase space is naturally determined by the set 
of thresholds and the corresponding symbols by the values of the Heaviside  
function $ H \left( \sigma (\x-T)\right)$.
With this dictionary in mind it is clear that the transducer in the RC case uses 
the phase space made available by the input signal to encode the message and, {\it vice--versa}, 
in the FC each node send to the others a code defined by its position in the available phase 
space. These are just two faces of the same coin.

\ms Our conclusion is that the existence of open designed dynamical modules leads to an 
extremely reach set of different dynamical behaviors making such units capable of carrying 
through various performances in response to different external stimuli.
But, on the other hand we have seen that in many cases the same functionality may be 
performed by different modules. The possible criteria for the selection of a particular 
module among other displaying the same function open to very interesting questions~\cite{S77}.

\ms The fine tune of the dynamics of a module by a complex input is a current subject of 
our efforts. As for further research in a connected direction, we mention the interesting 
problem of reverse engineering, control and bifurcations for such modules that we have just 
touched in Example~\ref{example3} . 

\ms  We are still far from a complete understanding of the way how module's architecture,  
parameters and inputs are related to functionalities in all cases, and 
{\it vice-versa}. Nevertheless we are convinced that our results give an insight in the 
fascinating interplay between modularity and dynamics.

\ms

\end{document}